\newif \ifcomments \commentstrue
\newif \ifsubmission \submissiontrue

\submissionfalse
\commentsfalse     

\newif \ifSP \SPtrue
\newif \ifCCS \CCSfalse

\ifsubmission
\newcommand{\sourceurl}{https://github.com/prof-anon}
\else
\newcommand{\sourceurl}{https://github.com/prof-project}
\fi

\ifCCS
\documentclass[sigconf, screen]{acmart}
    \renewcommand\footnotetextcopyrightpermission[1]{}
\fi

\ifSP
\documentclass[letterpaper,conference,compsoc]{IEEEtran}
\usepackage[noadjust]{cite}
\let\labelindent\relax
\usepackage[scaled=0.88]{helvet}

\fi

\usepackage[]{xcolor}
\usepackage{cancel}
\usepackage{soul}

\ifCCS
\else 
    \usepackage{amsmath}
    \usepackage[hyphens]{url}
    \usepackage{graphicx}
    \usepackage{hyperref}
\fi

\usepackage{booktabs}
\usepackage{makecell}
\usepackage{amssymb}
\usepackage{pifont}
\usepackage{enumitem}
\usepackage{tikz}
\usepackage{tabularx}

\newtheorem{remark}{Remark}
\newtheorem{definition}{Definition}
\newtheorem{requirement}{Requirement}

\newcommand*\circled[1]{\tikz[baseline=(char.base)]{\node[shape=circle,fill,inner sep=1pt] (char) {\footnotesize\textcolor{white}{#1}};}}

\definecolor{DarkGreen}{rgb}{0.0, 0.7, 0.0}
\definecolor{PastelOrange}{rgb}{1.0,0.5,0.0}

\ifcomments
    \newcommand{\kushal}[1]{\textsf{\color{orange}{[Kushal: #1]}}}
    \newcommand{\nerla}[1]{\textsf{\color{DarkGreen}{[nerla: #1]}}}
    \newcommand{\ari}[1]{\textsf{\color{red}{[Ari: #1]}}}
    \newcommand{\anote}[1]{\textsf{\color{magenta}{[AM: #1]}}}    
    \newcommand{\yan}[1]{\textsf{\color{cyan}{[Yan: #1]}}}
    \newcommand{\mahimna}[1]{\textsf{\color{purple}{[Mahimna: #1]}}}
    \newcommand{\ujval}[1]{\textsf{\color{violet}{[ujval: #1]}}}

    \newcommand{\todo}[1]{\textsf{\color{red}{[ToDo: #1]}}}
\else
     \newcommand{\kushal}[1]{}
     \newcommand{\ari}[1]{}
     \newcommand{\nerla}[1]{}
     \newcommand{\anote}[1]{}
    \newcommand{\yan}[1]{}
    \newcommand{\mahimna}[1]{}
    \newcommand{\ujval}[1]{}
     \newcommand{\todo}[1]{}
\fi

\newcommand{\mypara}[1]{\smallskip\noindent\textbf{#1}\hspace*{0.25em}}
\newcommand{\gasused}{\ensuremath{{g}}}
\newcommand{\block}{\ensuremath{{B}}}
\newcommand{\profbundle}{\ensuremath{\theta_{\text{PROF}}}}
\newcommand{\sharebundle}{\ensuremath{\theta_\text{Share}}}
\newcommand{\profblock}{\ensuremath{\mathcal{B}_\text{PROF}}}
\newcommand{\profshareblock}{\ensuremath{\mathcal{B}_\text{PROF-Share}}}
\newcommand{\winningpbs}{\ensuremath{\mathcal{B}^*_{\textsf {PBS}}}}
\newcommand{\pbsblock}{\ensuremath{\mathcal{B}_{\textsf {PBS}}}}
\newcommand{\earlyblock}{\ensuremath{\mathcal{B}_\text{Early}}}
\newcommand{\lateblock}{\ensuremath{\mathcal{B}_\text{Late}}}

\usepackage{subfig}
\usepackage{cleveref}

\begin{document}

\date{}

\title{PROF: \underline{Pr}otected \underline{O}rder \underline{F}low in a Profit-Seeking World}

\ifsubmission
    \author{}
\else
    \ifSP
        \author{
            \IEEEauthorblockN{
                Kushal Babel\IEEEauthorrefmark{2}\IEEEauthorrefmark{4},
                Nerla Jean-Louis\IEEEauthorrefmark{3}\IEEEauthorrefmark{4},
                Yan Ji\IEEEauthorrefmark{2}\IEEEauthorrefmark{4},
                Ujval Misra\IEEEauthorrefmark{6}\IEEEauthorrefmark{4},
                Mahimna Kelkar\IEEEauthorrefmark{2}\IEEEauthorrefmark{4},\\
                Kosala Yapa Mudiyanselage\IEEEauthorrefmark{5},
                Andrew Miller\IEEEauthorrefmark{3}\IEEEauthorrefmark{4},
                Ari Juels\IEEEauthorrefmark{2}\IEEEauthorrefmark{4}
            }\\
            \IEEEauthorblockA{
                \IEEEauthorrefmark{2}Cornell Tech,
                \IEEEauthorrefmark{3}UIUC,
                \IEEEauthorrefmark{6}UC Berkeley,
                \IEEEauthorrefmark{4}IC3,
                \IEEEauthorrefmark{5}Fidelity Center for Applied Technology
            }
        }
    \fi
    \ifCCS
        \author{Kushal Babel}
            \authornote{The first two authors contributed equally to this work.}
            \affiliation{
              \institution{Cornell Tech, IC3}
              \country{}
            }
            \author{Nerla Jean-Louis}
            \authornotemark[1]
            \affiliation{
              \institution{UIUC, IC3}
              \country{}
            }
            \author{Yan Ji}
            \affiliation{
              \institution{Cornell Tech, IC3}
              \country{}
            }
            \author{Ujval Misra}
            \affiliation{
              \institution{UC Berkeley, IC3}
              \country{}
            }
            \author{Mahimna Kelkar}
            \affiliation{
              \institution{Cornell Tech, IC3}
              \country{}
            }
            \author{Kosala Yapa Mudiyanselage}
            \affiliation{
              \institution{Fidelity Center for Applied Technology}
              \country{}
            }
            \author{Andrew Miller}
            \affiliation{
              \institution{UIUC, IC3}
              \country{}
            }
            \author{Ari Juels}
            \affiliation{
              \institution{Cornell Tech, IC3}
              \country{}
            }
    \fi
\fi

\ifCCS
    \begin{abstract}
Users of decentralized finance (DeFi) applications face significant risks from adversarial actions that manipulate the order of transactions to extract value from users. Such actions---an adversarial form of what is called maximal-extractable value (MEV)---impact both individual outcomes and the stability of the DeFi ecosystem. MEV exploitation, moreover, is being institutionalized through an architectural paradigm known Proposer-Builder Separation (PBS).

This work introduces a system called PROF (\underline{Pr}otected \underline{O}rder \underline{F}low) that is designed to limit harmful forms of MEV in existing PBS systems. PROF aims at this goal using two ideas. First, PROF imposes an ordering on a set (``bundle’’) of privately input transactions and enforces that ordering all the way through to block production—preventing transaction-order manipulation. Second, PROF creates bundles whose inclusion is profitable to block producers, thereby ensuring that bundles see timely inclusion in blocks.

PROF is backward-compatible, meaning that it works with existing and future PBS designs. 
PROF is also compatible with any desired algorithm for ordering transactions within a PROF bundle (e.g., first-come, first-serve, fee-based, etc.).
It executes efficiently, i.e., with low latency, and requires no additional trust assumptions among PBS entities.
We quantitatively and qualitatively analyze PROF’s incentive structure, and its utility to users compared with existing solutions. We also report on inclusion likelihood of PROF transactions, and concrete latency numbers through our end-to-end implementation.

\end{abstract} \maketitle
\else 
    \maketitle 
\fi

\pagestyle{plain}

\section{Introduction}

Decentralized finance (DeFi), meaning the ecosystem of financial applications on blockchains, has seen a dramatic rise in recent years.
At the time of writing, DeFi applications on Ethereum alone hold \$50 billion in value, and handle a daily transaction volume of \$2 billion~\cite{defillama}. In many common applications---decentralized exchanges and lending contracts to name a few---the \textit{order} in which transactions are executed critically impacts the outcome for users. 

Unfortunately, even in permissionless blockchains, the power to choose the transactions within a given block along with their ordering still lies with a single \textit{proposer} (miner or validator). This means an adversary, or even a profit-seeking entity could choose any ordering it desires, often at the expense of ordinary users.
The extent of profit achievable through such strategic reordering or selective censorship of user transactions is commonly captured under the umbrella of \textit{MEV} (\textit{Miner} or \textit{Maximal Extractable Value})~\cite{daian2020flashboys,babel2023clockwork}. This includes not just commonly-known strategies such as front-running and sandwich attacks but also a slew of highly sophisticated strategies~\cite{zihao2023demystifying,babel2023lanturn}. Conservative estimates put the value of MEV extracted on Ethereum at the expense of ordinary users at over \$500 million over 2 years~\cite{Qin2021QuantifyingBE}.

\mypara{Competing approaches to deal with MEV.}
Broadly speaking, two radically different approaches have emerged to deal generically\footnote{An extensive but separate line of work~\cite{bentov2017tesseract,ciampi2021fairmm,cowswap,ferreira2023credible} targets MEV from specific applications (such as AMMs), but in this paper, we focus our attention on generic blockchain-level approaches to deal with MEV.} with MEV.

The first, primarily academic, line of work has focused on \textit{eliminating} MEV (or at least minimizing it)---this has taken a number of forms with complementary guarantees. For instance, \textit{blind ordering}~\cite{cachin2001asyncbroadcast,malkhi2023mev,Kavousi2023blind,li2023transaction,li2023ratel} uses cryptographic techniques to keep transaction data hidden until the ordering is finalized. Separately, \textit{fair ordering}~\cite{cachin2022quick,kelkar2020order, zhang2020oligarchy,kiayias2024ordering} decentralizes the ordering across a committee instead of a single validator, and provides FCFS (first-come-first-served) ordering guarantees based on the time at which transactions are received by the committee. 
Despite providing strong guarantees against adversarial manipulation of the transaction ordering, this large body of literature has currently seen little success in practical deployments.
A significant obstacle is that deploying these protocols in their original form requires a complete revamp of critical parts of the system.
Validators of existing systems have not willingly adopted these protocols, as doing so would disregard their clearly evident profit-maximizing intentions.
 

On the other hand, protocols currently deployed in practice~\cite{flashbots, solana-jito, heimbach2023pbs} take a radically different approach.  Despite acknowledging the harmful effects of MEV, a common starting point for these protocols is the \textit{rejection of the honesty assumptions} commonplace in the literature attempting to minimize MEV. They argue that validators (a.k.a.~proposers) will always be profit-seeking entities, and making, e.g., honest-majority assumptions among them is a non-starter. With this viewpoint, they tend to accept MEV as is, instead attempting to only mitigate more catastrophic downstream effects---system instability and centralization. They facilitate all validators to take part in extraction \textit{equally} by selling their ordering power to more sophisticated actors who compete to build the most profitable block. 

In Ethereum, Proposer-Builder Separation (PBS) is the umbrella term for the idea of such a block-building marketplace and the widely used infrastructure to achieve it. Unfortunately, recent studies~\cite{weintraub2022flashbot, li2023flashbots} suggest that PBS has exacerbated the MEV problem for ordinary users.

\mypara{Recent middle-ground approaches.}
Recent industry efforts, such as MEV Blocker~\cite{mev-blocker}, MEV-Share~\cite{mevsharedocs}, and others~\cite{flashbots-protect,bloxroute-ethical} have tried to capture the best of both approaches---protecting users from harmful MEV extraction, while being able to integrate into existing systems with profit-seeking validators (albeit with additional trusted intermediaries).
Blocks built using these techniques, however, still have to compete with those built with more profitable approaches which do not aim to protect users.
In turn, user transactions will be included on the blockchain only in those rounds where the selected trusted intermediaries build the most-profitable block.
Furthermore, as transaction privacy increases to protect users of these systems, extracted MEV and thus the competitiveness of their blocks decreases.

This presents a new practical tradeoff, now between MEV protection and transaction inclusion rate.
Users may benefit from reduced MEV extraction, but at the cost of service degradation, as their transactions take longer to be included on chain.
Note that additionally, \textit{some} MEV is still often extracted from users in these systems, as economic benefits are not internalized among users but are rather leaked to arbitragers and validators (Section~\ref{sec:analysis-economics}).
Thanks to our insights in this work (summarized in Section~\ref{sec:salient}), we transcend the above limitations.

\begin{figure}
    \centering
    \includegraphics[width=\columnwidth]{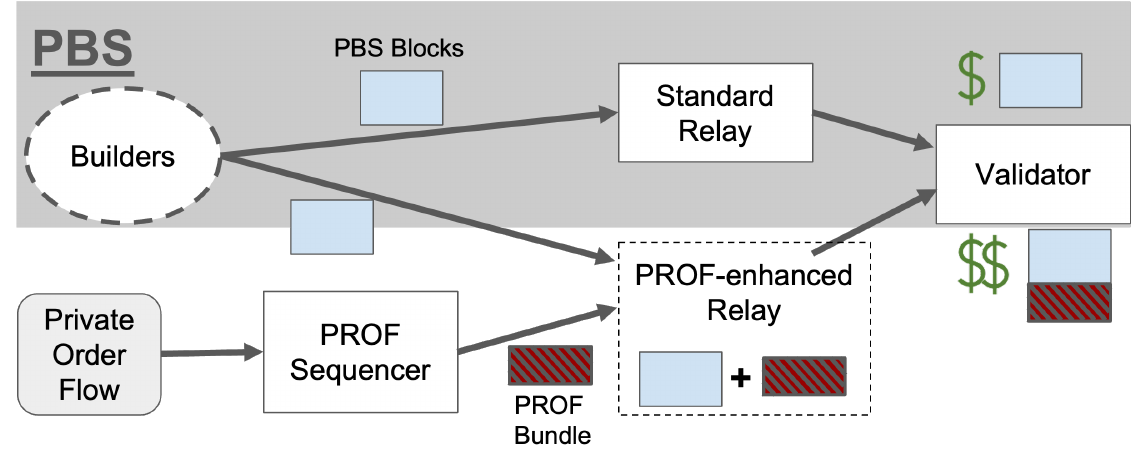}
    \caption{High-level design of PROF. Gray-shading shows existing PBS infrastructure. Without deployment of PROF, a block (shown in blue) is routed from PBS through a standard relay to a validator, resulting in validator profit {\color{green}{\$}}. With deployment of PROF, a PROF bundle may be added to a PBS-generated block to yield a new, enriched block with validator profit {\color{green}{\$\$}}. Note that a PROF-enabled relay here is simply a standard relay, but with additional logic to concurrently produce the enriched block.}
    \label{fig:overview}
\end{figure}

\mypara{PROF: MEV minimization in a profit-seeking world.}
In this work, we introduce PROF, a novel approach for dealing with MEV. In contrast to previous approaches, PROF achieves all three properties: It minimizes MEV extraction for users, it is able to integrate within existing systems with profit-seeking validators, and it does not degrade transaction inclusion-time. 

The key idea behind PROF is simple yet powerful. PROF does not take the  approach in existing systems of directly constructing full blocks that aim to provide the maximum profit to validators. Instead, PROF gives validators an \textit{additive take-it-or-leave-it} choice for a specific bundle (i.e., ordered set) of user transactions routed through PROF. A validator gets to choose whether to \textit{append} this bundle of
PROF transactions to the already crafted most-profitable block. 

Transactions handled by PROF are private---they are not sent to the public mempool.  By choosing to append PROF transactions, therefore, a validator obtains \textit{additional revenue} (from the transaction fees paid by PROF transactions) \textit{on top of the revenue from the best existing block}. In other words, PROF is incentive-compatible for (myopic) validators. We consider only myopic validators, i.e., validators that are optimizing only for the current slot, for two reasons: (1) The EIP-1559 transaction fee mechanism is itself not incentive compatible for non-myopic validators~\cite{roughgarden2021transaction}, and (2) 
If a user were deprived of inclusion through PROF in the current slot by a non-myopic validator, and was instead forced  to retry sending the transaction through the public mempool, the validator can earn a higher revenue by extracting value from the user.


\begin{definition}{\bf Incentive-Compatibility for Myopic Validators.}
    Consider a validator $V$ that has the choice of proposing any block from the set $\mathfrak{B}$ of valid blocks at blockchain state $s$. Let $\mathsf{revenue}(s, \block)$ denote the revenue earned by $V$ in block $\block$. 
    A mechanism is incentive-compatible for myopic validators if it selects the block $\text{argmax} _{\block \in \mathfrak{B}} \mathsf{revenue}(s,\block)$.
\end{definition}

As a direct consequence of its additive mechanism, PROF enjoys a high inclusion likelihood for its transactions since profit-seeking validators will choose the incrementally higher revenue of a PROF-appended block. At the same time, we emphasize that privacy for PROF transactions is maintained in the event that the validator leaves them out. 

PROF can use any desired ordering scheme for the transactions it handles, i.e., it is agnostic to ordering rules and techniques. This means that it can serve as a vehicle for any MEV minimization ordering protocol to be deployed in practical systems with profit-seeking actors. We briefly discuss in Section~\ref{sec:discussion} how PROF additionally can alleviate the main negative externality of an FCFS policy---a latency arms race similar to that observed with high-frequency trading~\cite{budish2015hftarm}.

We note that while our focus is on Ethereum in this paper, MEV landscape on some other blockchains (e.g., Solana), layer 2 rollups~\cite{timeboost-adoption}, and upcoming shared sequencing protocols (e.g., Espresso~\cite{espresso}) is similar to Ethereum, and insights from PROF could feasibly be applied there.


\begin{table*}
\centering
\begin{tabular}{|c|c|c|c|c|}

\hline
{\bf \makecell{Protocol opted into\\ by a validator}} & {\bf \makecell{\circled{1} MEV Protection \\ for Users}} & {\bf\makecell{\circled{2} Validator\\Incentive-Compatibility}} & 
{\bf \makecell{\circled{3} Impact on\\ Inclusion-Time}} & {\bf \makecell{\circled{4} Trust Assumptions}} \\[1.5ex]
\hline
\rule{0pt}{2ex} \makecell{Temporal Fair Ordering} & \makecell{Through Joint \\Sequencing} & \color{red}{\ding{55}} & \color{DarkGreen}{None} & Majority Honest Committee \\ \hline
\rule{0pt}{2ex} \makecell{Blind Ordering} & \makecell{Through Privacy} & \color{red}{\ding{55}} & \color{DarkGreen}{None} & \makecell{Majority Honest Committee \\ or TEE} \\ \hline
PBS (MEV-Boost) & \color{red}{\ding{55}} & \color{DarkGreen}{\checkmark} & \color{DarkGreen}{None} & Trusted PBS Relay \\ \hline
MEV-Blocker / MEV-Share & {\color{red} Privacy Tradeoff with} \circled{3} &  \color{DarkGreen}{\checkmark} & {\color{red}Tradeoff with} \circled{1} & \makecell{Trusted PBS Relay + \\  Trusted MEV-Blocker / \\ MEV-Share Node + \\ Trusted Builders and Searchers} \\\hline
\makecell{PROF} & \makecell{As Enforced by Sequencer \\ (TEE-Enforced Privacy \\ in Our Implementation) } & \color{DarkGreen}{\checkmark} & \color{DarkGreen}{Minimal}  & Trusted PBS Relay \\
\hline
\end{tabular}
\caption{
    PROF achieves the properties of: (1) MEV Protection for Users, by keeping the transactions private until they are committed to by the proposing validator; (2) Validator Incentive-Compatibility, i.e., rational (profit-maximizing) validators are always given the most profitable block; and (3) Minimal Impact on Inclusion-Time, i.e. the PROF bundle is included in the next block with a high probability (see Section~\ref{section:fee-latency}). PROF achieves these properties (4) without introducing any additional Trust Assumptions besides the existing trusted relay, and without introducing a tradeoff between MEV Protection for Users, and Inclusion-Time of their transactions.
    Fair ordering, blind ordering, or other fair-ordering approaches can be incorporated into PROF sequencer. 
    }\label{tab:sys-comparison}
\end{table*}

\subsection{PROF Overview and Contributions}
\noindent \textbf{Background.}
Although a validator has complete authority to choose and order transactions in its block, the complexity of actually finding the most profitable ordering has led to a division of labor on Ethereum. Validators \textit{auction} off their block-construction right to the highest bidder. Sophisticated actors---\textit{builders}---who specialize in MEV extraction compete in this marketplace, whose general design is referred to as  Proposer-Builder Separation (PBS).

The current instantiation of PBS on Ethereum is the MEV-Boost system~\cite{mevboostdocs}, which exists \textit{outside} the core Ethereum protocol\footnote{While PBS could feasibly be instantiated as part of the core Ethereum protocol (dubbed \textit{enshrined}-PBS), this poses several challenges---many of which are active research questions; see Section~\ref{sec:related}.}. The permissionless nature of the marketplace necessitates the use of a third entity in MEV-Boost---a \textit{trusted relay}---to act as a bridge between proposers and builders.

Relays serve two goals. They prevent both: (1) A malicious proposer from ``stealing'' a builder's MEV strategy and private transactions and (2) Malicious builders from submitting invalid blocks or bids. In principle, relays could use trusted execution environments (TEE) to simulate a builders' blocks and check validity without learning their contents. But even then, relays still need to be trusted to deliver the block to the validator after obtaining a commitment.

Relays play a pivotal role in the design for PROF. In fact, relays allow PROF to integrate into existing PBS infrastructure \textit{without making any additional trust assumptions}.

Figure~\ref{fig:overview} illustrates the high-level design of PROF and its interaction with PBS, specifically its integration with relays. 

Table~\ref{tab:sys-comparison} compares the properties of PROF with those of existing approaches. We emphasize that the relative simplicity of our design also translates to avoiding the trust assumptions made in the existing middle ground approaches like MEV-Blocker and MEV-Share; these trust assumptions have already created significant barriers to decentralization in practice~\cite{yang2024decentralization}.

\mypara{Basic PROF design (Section~\ref{sec:design}).}
PROF consists of two components: the \textit{sequencer} and the \textit{bundle merger} 

The PROF sequencer ingests user transactions, and sequences them into a \textit{bundle}, according to any pre-specified rule $\mathsf{R}$. As mentioned before, the PROF design is agnostic to which sequencing rule is used. For instance, $\mathsf{R}$ could represent a ``fair ordering'' (decentralized FCFS) policy~\cite{kelkar2020order} (see Section~\ref{sec:discussion}).

The PROF merger resides by default at a PBS relay\footnote{For an optimistic PBS relay, i.e., a relay that does not simulate the block from builders, the PROF merger resides at the builder itself. See Section~\ref{sec:buildermerge}.} (PROF bundle merger instances can operate independently at multiple PBS relays, but in our explanation and illustrations, we focus on a single relay). It performs the following simple task: It takes the winning PBS block $\winningpbs$, and appends to it the transaction bundle provided by the PROF sequencer. It thereby creates a new block $\profblock$ to be sent to the proposer. We note that blocks are seldom full; this is due to the nature of Ethereum's EIP-1559 fee mechanism~\cite{roughgarden2021transaction}---this allows the PROF bundle to be appended without hitting the block-size limit. 

PROF's additive take-it-or-leave-it property is realized through two key observations: (1) Merging the PROF bundle at the \textit{relay} enables us to \textit{supplement} the most profitable block from the winning builder instead of competing with it. This observation is extremely simple, yet overlooked in all previous designs; (2) To the extent that the transactions are exclusive (i.e., transactions not present in the public mempool) to PROF, we can provide additional revenue in the PROF-enriched block simply from user transaction fees. To maintain transaction privacy for user transactions, we have designed PROF to run inside a TEE. (This is not strictly necessary for compatibility with PBS infrastructure or trust models. Today relays are fully trusted in practice.)

\mypara{PROF-Share: An enhanced design (Section~\ref{section:double-auction-design}).}
A recent protocol, MEV-Share~\cite{mevsharedocs}, aims to protect users not by minimizing MEV, but instead by requiring that a portion of the extracted MEV is \textit{returned to the user}.
MEV-Share currently requires trust in the operator.
The operator requires that any revenue obtained through \textit{back-running} of a transaction bundle, meaning extraction of MEV from the transaction bundle by executing an arbitraging transaction \textit{immediately afterward}, is shared with users in the transaction bundle.

Inspired by this design, we propose an enhanced version of PROF called PROF-Share. 
After transactions are sequenced, PROF-Share includes back-running transactions at the end, and returns the resulting MEV profits back to users.
In general, PROF-Share provides even better transaction execution, i.e., trade execution pricing, for users than basic PROF.
PROF-Share users also generally enjoy economic outcomes and privacy superior to MEV-Share. We summarize these benefits in Section~\ref{sec:salient}. 

\mypara{Practical analysis (Section~\ref{sec:analysis} and Section~\ref{section:fee-latency}).}
We provide a rigorous analysis of PROF, and a comparison to other mechanisms (such as (vanilla) PBS and MEV-Share) on three axes: (1) Utility (specifically, effective trade execution price) for users; (2) Incentive-compatibility for validators; and (3) Inclusion-time for user transactions.
We utilize real-world data for the Ethereum PBS auctions to demonstrate that the latency imposed by PROF has minimal effect on transaction inclusion-time.
Moreover, this effect of latency is simply offset by any transaction fee that users pay on top of the mandatory base fee.
For instance, for a latency of 10ms imposed by a PROF bundle of size 750k gas (roughly 5 AMM swap transactions),
and a transaction fee overhead of 10\% (as a percentage of the base fee), the proposer in $>$95\% of PBS auctions gains more revenue by choosing the block with the PROF bundle included, than merely choosing the winning auction block without the PROF bundle.
Therefore a rational proposer will include the PROF bundle with a high probability.
In the remaining cases where the PROF-enriched block is not competitive due to the latency involved in its production, the proposer receives the best block, i.e., the winning block from the PBS auction. Thus a proposer never loses anything as a result of having the option of choosing PROF. Finally, we note that if the PROF bundle isn't included in the current block, PROF transactions will just be deferred until the next block (or, rarely, a later block).

\mypara{Implementation (Section~\ref{sec:impl}).} We provide an end-to-end implementation of PROF along with latency benchmarks. 
Our benchmarks show that in Ethereum, PROF imposes a base latency of only 6.25ms, with additional latency of $<$1ms per transaction in a PROF bundle\footnote{Latency depends on transaction types, and is more accurately characterized by the gas used by a transaction (see Section~\ref{sec:eval-latency})}.
This demonstrates that PROF can be practically deployed on Ethereum today. While relays are currently fully trusted in the Ethereum PBS ecosystem, we still run our PROF implementation inside trusted hardware to provide defense in depth. As a result, our latency benchmarks are conservative even for practical deployment in the wild.


\subsection{Summary of Key Insights and Properties}
\label{sec:salient}
PROF transactions enjoy a strong chain of custody while participating in the PBS ecosystem: They are kept private until the proposer commits to including them.
This key insight leads to the following properties:
\begin{enumerate}
\item PROF provides incremental revenue on top of the winning PBS block, making it incentive compatible for the proposer to include protected transactions. 
\item PROF transactions bypass the competition in PBS,
thus enjoying high inclusion likelihood for minimal fee. Consequently, PROF-Share users, unlike MEV-Share users, get to keep all of the backrunning profits rather than sharing it with validators.
\item PROF-Share safely releases entire transaction contents to arbitragers rather than partial leakage in MEV-Share. This allows for efficient backrunning while affording privacy to the user transactions.
\item Unlike other block-building mechanisms that attempt to provide privacy to users~\cite{mevsharedocs, SUAVEspecs:2024, mev-blocker}, PROF mitigates against any grinding attacks based on the information leaked in form of proposer's revenue.
\item PROF-Share, unlike MEV-Share, ensures that all its users are backrun together. This allows users to organically backrun each other and prevents leakage of value from users to backrunning arbitragers\footnote{We recommend that existing redistributive mechanisms such as MEV-Share and MEV-Blocker should also explore the possibility of banding together as many user transactions as possible before opening them up to backrunning by arbitragers.}.

\end{enumerate}

\section{Background}

\subsection{Offchain Proposer-Builder Separation (PBS)}
\begin{figure}
    \centering
    \includegraphics[width=.99\columnwidth]{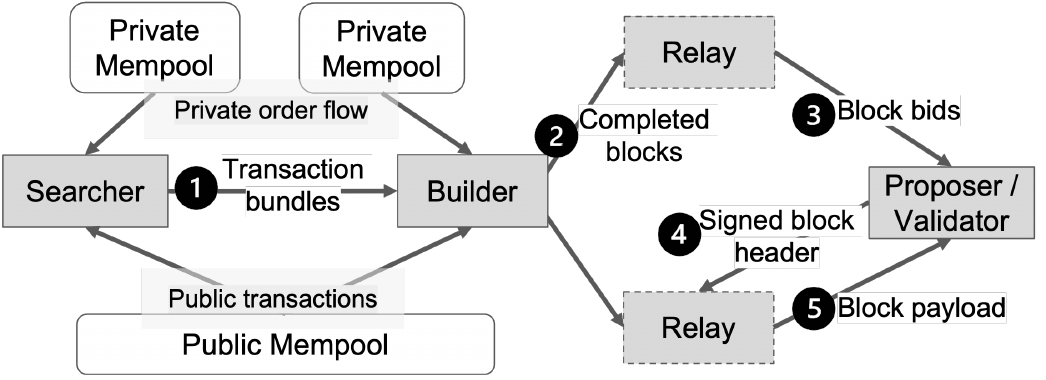}
    \caption{MEV-Boost design. ``Searchers'' and ``builders'' together create transactions bundles from their private and public mempool. Builders submit a complete block to a PBS relay, which then forwards the block header, along with the bid, to the proposing validator. The validator receives the corresponding complete block after it commits to the block, by signing the block header.
    }
    \label{fig:mevboost}
    
\end{figure}

As mentioned earlier, PBS refers to the idea of validators / proposers auctioning off their ordering right to sophisticated \textit{builders} who specialize in MEV extraction by choosing transactions and their ordering in the block.

\mypara{MEV-Boost.}
Although adding PBS to the core Ethereum protocol is a topic of active research, a version of PBS---MEV-Boost~\cite{mevboostdocs} has already been deployed \textit{external to the Ethereum protocol} since 2022. MEV-Boost patches the standard validator client to accept blocks from an external builder marketplace. An estimated 85-90\% of validators run MEV-Boost alongside the vanilla Ethereum client~\cite{heimbach2023ethereums}. We illustrate the workflow of MEV-Boost in Figure \ref{fig:mevboost}.

Block construction begins with \textit{searchers}, specialized actors who identify profit opportunities by sequencing collections of transactions into \textit{bundles}. Searchers forward bundles to block \textit{builders} for use in the construction of blocks.

\textit{Relays} exist as trusted intermediaries between the builders and the current proposer/validator. Recall that PBS realizes an \textit{auction} for blocks. In MEV-Boost, builders submit their blocks to relays. The bid for each block is calculated as the value (in terms of the native currency ETH) accrued to the current proposer's address. Each relay  simulates incoming builders' blocks to ensure their validity and then provides the maximum bid, along with the corresponding block header, to the proposer. A proposer may continuously poll relays to take advantage of any rise in bids as new blocks are formed by builders. 

A proposer may choose to connect to multiple relays. After choosing the maximum bid it receives from a relay, the proposer signs the header of the corresponding block, and sends it back to the relay to request the full block. The proposer's signature \textit{commits} it to a block before it sees the block contents, in the following sense. A proposer that signs and proposes a different block---and has thus signed two distinct blocks at the same height---is subject to Ethereum's existing slashing mechanism and will incur a financial penalty~\cite{ethslash}. Looking ahead, PROF utilizes commitments from proposers to give protection to user transactions.

A relay needs to be trusted by both builders and proposers. Builders trust the relay to not release their blocks in any other circumstance, and run the auction with integrity.
On the other hand, proposers trust the relay to ensure that the winning block is valid and will be released in a timely. There is an optimistic variation of the basic relay that does not perform validation in order to cut down on latency, and relies on a reputation / bond of the participating builders~\cite{optimistic-relay-doc}. We show how PROF can also take advantage of the lower latency of the optimistic relay in Section \ref{sec:buildermerge}.

\subsection{MEV-Share}\label{section:mevsharebackground}
MEV-Share~\cite{mevsharedocs}  is a deployed service that allows users to selectively share data about their transactions with searchers, who bid to include the transactions (via bundles) in the blocks eventually built by the builders. The users are promised profits from any arbitrage opportunities created by their transaction would be shared with them.

Users first send their transactions to MEV-Share Matchmaker Node, which is a trusted intermediary.
According to users' privacy preference, the Matchmaker shares information about received transactions, known as \textit{hints}, to the searchers. These hints can include logs with information on what trading pools the user is accessing, the smart contract and function being called by the transaction, or just the transaction's hash. 

The searchers submit bundle templates back to the Matchmaker in order to attempt to extract MEV from private transactions (currently via backrunning, i.e., follow-on transactions). These templates include spaces for private-transaction inclusion. The specific transaction to insert can be pre-specified in the template by the searcher or assigned by the Matchmaker. The bundles are simulated by the Matchmaker to see which searcher extracted the most MEV from the private transaction with their bundle. The searcher distributes the extracted MEV among itself, the user, the builder, and the validator / proposer. 
The bundle that accrues the most value to the builder is forwarded to the builder.

By default, 90\% of the estimated MEV generated by these transactions or bundles is required to be redistributed to the users, but this can be adjusted by users. 
MEV-Share also allows users to send a portion of the MEV to proposers, to potentially get faster inclusion. Users are also instructed to set a priority fee so their transaction can be included if it is not bid on and included in a searcher bundle.

Note that this mechanism has a tradeoff between the inclusion-time of a transaction and the privacy it can demand from the market participants. The more information a transaction reveals, the more lucrative it gets for builders and validators to work to finalize it, as it allows them to extract more MEV profits.

\subsection{EIP-1559: Dynamic Base Fee Mechanism}
The Ethereum Improvement Proposal (EIP) 1559 is the currently deployed transaction-fee mechanism in Ethereum. It aims to enhance usability through a dynamic congestion control mechanism over transaction fees, and at incentive compatibility for both users and miners by discouraging off-chain collusion ~\cite{roughgarden2021transaction}. EIP-1559 replaced the fee-auction mechanism previously used for transaction inclusion with a new mechanism that uses changes in block sizes to signal congestion or demand. Under EIP-1559, users must pay a dynamic but deterministic base fee for inclusion in the new block, and can optionally add a ``tip'' to incentivize inclusion, especially in times of congestion.
The base fee is ``burned,'' meaning removed from circulation by the protocol. 
Tips, however, accrue to a designated address in a block called the ``coinbase.''


As a characteristic of the dynamic base fees mechanism,  blocks are generally only partially full. EIP-1559 is designed to keep blocks 50\%-full on expectation.

We exploit this property in the design of PROF, as it ensures that there is often empty space available in the block, and PROF transactions that pay any non-zero tips provide incremental revenue to a proposer / validator. 

\subsection{Trusted Execution Environments}

Trusted Execution Environments (TEEs)~\cite{Anati2013InnovativeTF, Costan2016IntelSE, McKeen2013InnovativeIA, Hoekstra2013UsingII} are tools that enable clients to outsource computations to a remote, untrusted server. A TEE operates in isolation, ensuring that even the host operating system cannot tamper with or access the memory within the TEE. Clients use remote attestation to receive integrity guarantees through a signed attestation report from the hardware manufacturer, confirming that the server is running the correct code on authentic hardware.

Due to their capability to enable both performant and secure computations, TEEs have attracted significant interest, particularly in the untrusted environments of decentralized applications. Use cases include private block-building ~\cite{flashbotsSGXbuilder}, order flow auctions \cite{flashbotsMEVMSUAVE}, private smart contracts \cite{cheng2019ekiden}, oracles for off-chain data \cite{zhang2016towncrier}, etc.

\section{Design}\label{sec:design}
\begin{figure*}
    \centering
    \includegraphics[width=.8\textwidth]{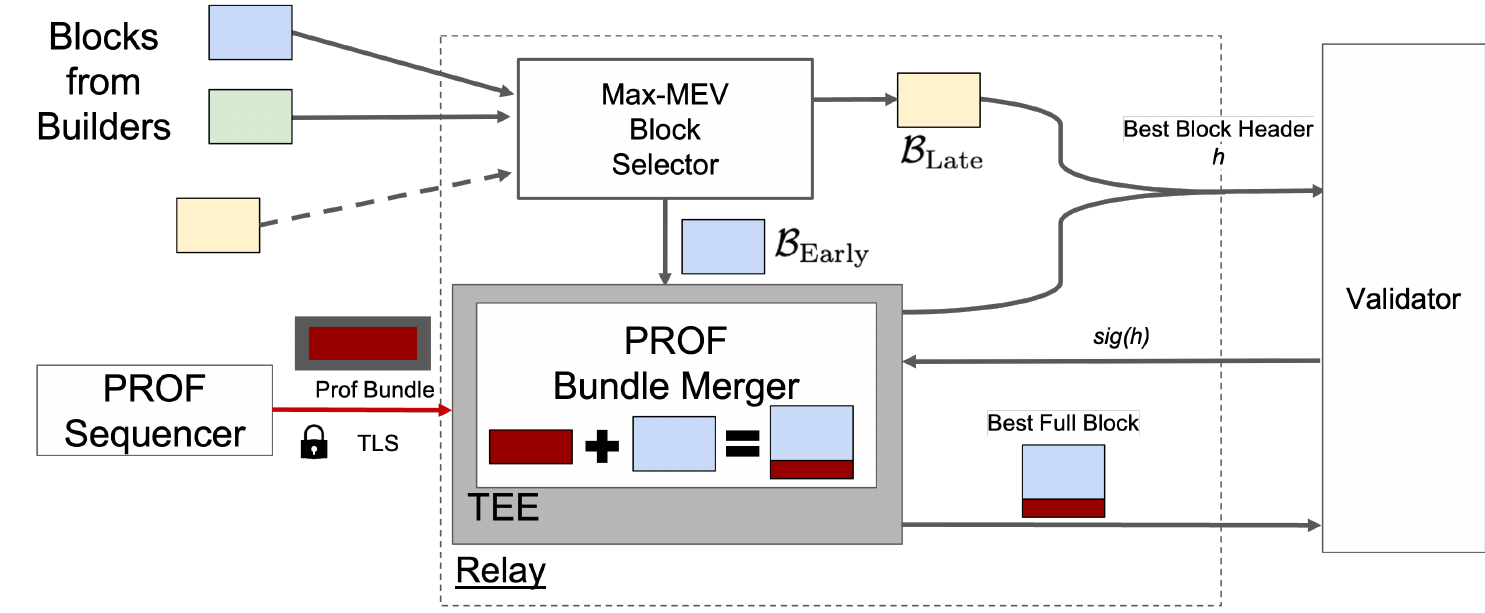}
    \caption{PROF design. We emphasize that the PROF sequencer here is a black-box and could be a decentralized protocol such as~\cite{kelkar2023themis}. While relays are  already trusted intermediaries in PBS, to achieve defence in depth, we implement the PROF bundle merger inside a TEE as a defence (see Section~\ref{subsec:tee-depth})
    }
    \label{fig:design}
\end{figure*}

\noindent We first recall the key system goals of PROF and briefly explain the key aspects of our design that help us achieve those goals.

\begin{itemize}
    \item \textit{Take-it-or-leave-it choice for proposers}: PROF only  supplies two kinds of blocks (and block headers) to any proposer: Either a block that has PROF transactions included in order, or a block from the PBS system that has no knowledge about the PROF transactions.
    \item \textit{Privacy of PROF transactions until block commitment}: PROF reveals the contents of PROF transactions \textit{only after} a proposer has committed to the block containing them (through a signed header). 
    \item \textit{Incentive compatibility for profit-seeking proposers}: All else being equal, a PROF-enabled relay never supplies a lower-value block to a proposer than a vanilla relay without PROF. This is achieved by constructing a PROF-enriched block concurrently with a winning PBS block, and presenting only the better of the two.
    \item \textit{Backward compatibility}: PROF requires no change to either proposers (validators) or builders in the PBS architecture. Moreover, the trust model for the validators remains the same: They trust relays to provide headers of only valid blocks and release block contents in response to signed block headers.
    \item \textit{Support for any transaction-ordering policy}: PROF abstracts out its transaction-ordering component into a black-box logical component called the PROF sequencer. PROF is agnostic to the sequencer's implemented transaction-ordering policy. 
\end{itemize}

Figure~\ref{fig:design} depicts the concrete logical design of PROF. 

In what follows, we give the details of our design of the PROF sequencer, and the PROF bundle merger, followed by a detailed discussion of the threat model and backward compatibility of our design.

\subsection{PROF Sequencer}
The PROF sequencer is responsible for collecting transactions from a protected (encrypted) order flow into an ordered sequence of transactions known as a bundle. While these transactions could also appear in the public mempool, and / or in the PBS supply chain, transactions that are sent exclusively to the PROF sequencer enjoy protection, while increasing the likelihood of the inclusion of the PROF bundle in the next block.
PROF sequencer then sends this bundle to the PROF bundle merger. 

The sequencer can implement any transaction-ordering policy, from fair ordering policies~\cite{kelkar2020order} to policies that prioritize transactions based on fees.
Crucially, this policy is not allowed to take into account the details of the contents of transactions themselves (except for the gas price).

Transactions are sent to to the sequencer over a secure channel (TLS connection), so that the sequencer may be viewed as operating on an encrypted mempool. The sequencer enforces transaction confidentiality during the bundling process. In our implementation, the sequencer is instantiated in a TEE, but in principle could be realized alternatively via MPC. 

Transactions that appear valid on their own (at the state obtained after executing the last finalized block) can turn out to be invalid depending on the position at which they are included in the next block. A transaction $T$ can be invalidated due to a few reasons: a another transaction $T'$ ordered before $T$ can consume all the balance meant for paying the transaction fees of $T$, $T'$ has the same ``nonce'' (sender's sequence number), etc.
As PROF transactions are appended to the PBS block obtained from the auction, the PROF sequencer can never be sure beforehand whether a given transaction is going to be valid when included in the final block. 
The PROF sequencer nevertheless checks validity in a best-effort manner by assuming that the transactions in the PROF bundle are executed on top of the last finalized block. 
Looking ahead, PROF bundle merger removes any invalidated transactions before creating the final block. This includes duplicate transactions, i.e., any transactions that also appear in the winning block from the PBS auction.

\subsection{PROF Bundle Merger}\label{section:single-auction-design} 
\mypara{Overview.}
The PROF bundle merger, hosted at the PBS relay, takes a PROF bundle from the sequencer and appends it to the end of the winning PBS block.
While PBS relays are already trusted intermediaries, we host the PROF bundle merger inside a TEE as defence in depth. We discuss in detail more reasons behind this choice, along with some caveats, in Section~\ref{subsec:tee-depth}.

The role of a PBS relay is to facilitate a first-price auction for builders' blocks in order to produce the maximal rewarding block for the proposer.

Note that the relay does not reveal a block's transactions until the proposer commits (via block-header signing)\footnote{If the validator violates its commitment by proposing another block header to the network, its stake gets slashed by the blockchain protocol due to equivocation.} to proposing the block offered by the relay.
Therefore, PROF gives the validator a choice: Take the protected PROF bundle as is, or don't take it all.
Since the PROF bundle provides additional revenue to the proposer (through transaction fees, and an optional direct payment), appending the PROF bundle to the PBS block only increases the reward to the proposer.
A PROF-enabled relay is thus strictly more lucrative to the proposer than a vanilla PBS relay, all else being equal.
Note that the PROF transactions do not compete against the builders' blocks, but rather merge with the most profitable block and provide incremental revenue to the validator. 
Hence, PROF transactions do not suffer degradation in their chances of inclusion for this reason.




\mypara{Latency consideration.} While a PROF bundle provides additional revenue to the proposer and protects users against ordering-based exploits, the PROF bundle merger does impose an additional latency $\delta$.
PROF incurs this extra latency $\delta$ because the PROF bundle must be simulated (at the bottom of the winning PBS block) in order to ensure that the PROF-enriched block is valid. 

For the proposer to get the header response from a PROF-enabled relay at the same time as that from a relay running without PROF, the PROF-enabled relay has to start the bundle merging process in advance. 
A parameter $T$ in PROF denotes the time instant at which the PROF bundle merger starts the merging process, by taking in the winning PBS block so far.
The PROF-enriched block is available at time $T+\delta$.
As we detail below, we design the PROF bundle merger so that the relay can continue running the MEV auction (beyond time $T$) concurrently with PROF simulation, mitigating any impact of latency on the profitability (and thus, incentive compatibility) of the proposer. We analyze in detail the impact of this latency on the inclusion chances of PROF transactions in Section~\ref{section:fee-latency}.

\mypara{Details.} After the PROF sequencer collects user transactions in a bundle, the bundle is sent to the PROF bundle merger using a secure TLS connection for inclusion in the next possible block. The bundle merger continuously listens for newer bundles from the sequencer, and replaces the previously communicated bundle with the new one.
Concurrently, the relay continues its normal operation of collecting blocks from builders and determining which block has the highest bid for the slot. Note that the bid for a block is simply the value that accrues to the proposer's address as a result of applying the block on the current state.

The proposer continually queries the relay for the highest bid (and the associated block header) among the builders, using the \verb|getHeader| API call. Note that the relay usually calculates this bid by simulating the builders' blocks and computing the difference in the proposer's balance before and after the simulation. (Some relays optimistically compute this bid by forgoing the simulation check, and directly trusting the builders to send the bid value.)
It is important to note that the proposer is free to submit \verb|getHeader| requests to many relays (including non-PROF relays).
When the proposer is satisfied by a bid, it commits to the bid by signing the corresponding block header.

At time $T$ (a parameter of PROF), the relay inputs the block (denoted $\earlyblock$ in Figure \ref{fig:design}) having the highest bid (up until time $T$) to the PROF bundle merger.
The PROF bundle merger simulates the PROF bundle at the end of the input block, i.e. the winning block from the MEV auction in PBS. Transactions in the PROF bundle that are duplicate or invalid during simulation are discarded from the bundle. The PROF bundle merger outputs only the header of the combined block and the total revenue to the proposer, which we call the PROF-enriched bid. The choice of $T$ is dictated by the latency $\delta$ of the bundle merger implementation, and we discuss it in detail in Section~\ref{section:fee-latency}.

While the bundle merger is simulating the PROF bundle, the PROF relay continues accepting bids from builders concurrently. 
When the bundle merger reveals its output, the relay compares the PROF-enriched bid and any late bids (denoted $\lateblock$ in Figure \ref{fig:design}) from the builders that arrived after time $T$.
The header of the highest-value block, along with the proposer's revenue therein, is returned to the proposer.
This concurrent simulation of a PROF bundle alongside the MEV auction in PBS ensures that the proposer does not miss out on any late blocks in the auction that alone have a higher revenue than the PROF-enriched block.
As a result, concurrent bidding preserves the property that PROF always results in proposer revenue at least as high as any PBS-generated block. 

Assuming the proposer chooses a PROF-enriched block, only after the proposer signs and sends to the PROF relay a header on it, committing the proposer to this block, does the PROF bundle merger return the contents of the final block (containing the PROF bundle) to the proposer.

\subsection{Other Design Considerations}
\label{sec:design-considerations}
\mypara{Transaction Fee flows from PROF users to proposers.} In PBS, builders usually set their own addresses as the \texttt {fee\_recipient} (also known as ``coinbase'') field in blocks they build to collect transaction fees and any MEV rewards from searchers that pay to the coinbase.
To pay the proposer, they then include an explicit payment transaction in the block.
This norm does not suit PROF well, because transaction fees in PROF transactions flow to the coinbase, and thus to the builder's address, rather than the proposer's.

To route PROF transaction fees instead to proposers, one solution is to use the ERC-4337 \emph{Account Abstraction} (AA) standard~\cite{eip4337}.
ERC-4337 enables users to have their accounts implemented in the form of smart contracts with customizable authentication methods, rather than just initiating transactions via an externally owned account (EOA) with a signing key pair, as is typical today.
In this standard, users create a pseudo-transaction object called a \texttt {UserOperation} and send to a special entity called an AA bundler (not to be confused with bundles in PBS).
The AA bundler collects \texttt{UserOperation} objects and bundles them into one EOA transaction.
A special smart contract called \texttt{EntryPoint} can take in AA bundles and process  \texttt{UserOperation}s one by one.
With ERC-4337, PROF users can send their transactions in the form of \texttt{UserOperation} objects, and the PROF bundle merger bundles the valid ones into an EOA transaction.
The logic of the \texttt{EntryPoint} contract is customizable, and allows for PROF transaction fees to be credited to the proposer's address.
In the context of PROF, the bundle merger could act as the AA bundler, and the PROF users would need to set up and use individual AA accounts rather than EOA accounts. We emphasize that this is only needed when builders do not set the proposer's address as the coinbase.


\bigskip
\noindent{\bf Bundle-merging at builders and optimistic relaying.} 
In Section~\ref{sec:buildermerge}, we propose a hybrid deployment model for PROF, where the PROF bundle merger resides at builders rather than relays.  The hybrid model takes advantage of the increasingly popular optimistic relays in order to dramatically cut down latency at the relay, and makes minimal changes to a PBS relay for enablement of PROF.

\begin{figure}
    \centering
    \includegraphics[width=.8\linewidth]{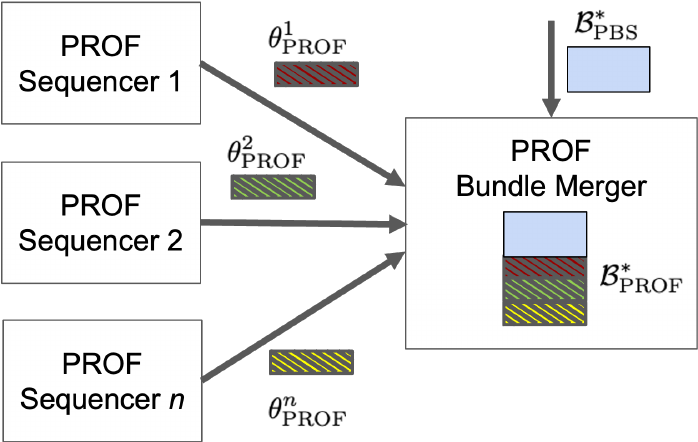}
    \caption{Protected bundles from multiple PROF sequencers can be included in the final PROF-enriched block.}
    \label{fig:sequential-sequencers}
\end{figure}

\bigskip
\noindent{\bf Multiple PROF Sequencers.}
\begin{figure}
    \centering
    \includegraphics[width=.8\linewidth]{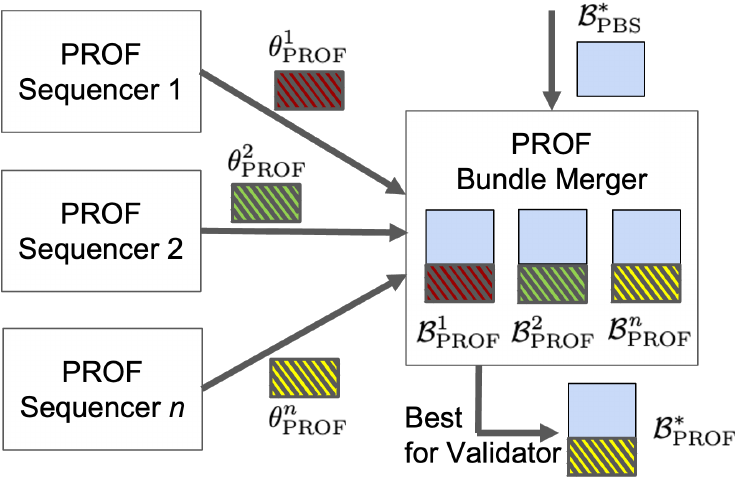}
    \caption{PROF allows for multiple sequencers to operate concurrently, and chooses the best PROF-enriched block for the validator. }
    \label{fig:concurrent-sequencers}
\end{figure}
While our protocol description has focused on one PROF sequencer (the sequencer black-box can itself be implemented through a decentralized fair and / or blind ordering protocol), PROF can work with multiple PROF sequencers, each of which could be implementing a different ordering policy.
Consider $n$ different PROF sequencers, each producing $\profbundle^1, \profbundle^2,...,\profbundle^n$ respectively. 
As shown in Figure~\ref{fig:concurrent-sequencers}, the PROF bundle merger simply merges each $\profbundle^i$ concurrently and independently with the winning PBS block $\winningpbs$ to obtain PROF-enriched blocks $\profblock^i (i = 1,...,n)$.
The protocol then simply selects the best PROF-enriched block $\profblock^*$, as measured by the revenue accrued to the proposer.
However, in doing so, all the other $n-1$ PROF sequencers' bundles get left out of the final block.
To overcome this, the PROF bundle merger could append to $\winningpbs$ all the bundles $\profbundle^1, \profbundle^2,...,\profbundle^n$ (according to some ordering policy, like FCFS, size of bundle, etc.) to obtain a single PROF-enriched block, that is at least as valuable as $\profblock^*$ to the proposer (Figure~\ref{fig:sequential-sequencers}).
The two approaches are on a spectrum of parallel v/s sequential merging (and thus latency), and revenue accrued to the proposer. The concrete choice of how many PROF bundles are merged together to obtain the final PROF-enriched block would depend on the parallel computation resources available, and the latency of merging.

\subsection{Threat Model and Backwards Compatibility}\label{section:threat-model-single}
PROF leaves the threat model and workflow for the builders in PBS unchanged.
As we explain below, relay operators need not embrace any new trust assumptions beyond those currently in the PBS architecture.
In line with the existing threat model for PBS: (1) PROF presumes that proposers and relays do not collude.
Specifically, relays report equivocations by malicious proposers, advancing evidence in the form of multiple signed block headers for the same slot.
This ensures that a malicious validator is cryptoeconomically committed to including the PROF transactions once they are revealed to it;
(2) PROF presumes that relays are not actively malicious. Relays release the full contents of blocks to the network in response to signed headers from proposers. 
Thus, once a block is committed to by a proposer, PROF transactions are made available to the entire network (and not exclusively to the operator of the relay).

While assuming that relays are not actively malicious, so long as the TEE for PROF operates correctly, PROF ensures confidentiality of PROF transactions from an honest-but-curious (passively adversarial) relay---until a proposer has committed to them. PROF's use of a TEE provides stronger security in this sense than existing PBS infrastructure which requires builders to trust relays not to prematurely disclose or exploit transactions in their submitted blocks.

\mypara{The Case for TEEs.}\label{subsec:tee-depth}
The PBS architecture today includes a number of strong trust assumptions, such as the searcher trusting the builder not to steal transaction bundles, and the relay acting as a trusted intermediary between the builder and validator. 
A PBS relay is trusted by builders to not leak transactions. 
Our use of a TEE at a relay in PROF, then, represents a  \textit{defense in depth} approach that confers a \textit{stronger} confidentiality model for PROF transactions than PBS alone does for transactions in an ordinary relay. It is similar
to other efforts in the industry, such as the Flashbots SGX Builder\footnote{\url{https://github.com/flashbots/geth-sgx-gramine}}.
While vulnerabilities in TEEs such as Intel SGX and side channel leakage\footnote{Note that as we only allow limited number of simulations of PROF transactions (see Section~\ref{subsec:mitigating-attacks}), any leakage from side channels during simulations in a TEE is already mitigated to a certain extent.} could compromise privacy guarantees~\cite{schaik2022sgxfail,zhang2020sidechannel}, such failure is not catastrophic, but would merely mean a fallback to the current PBS trust model. In short, the use of a TEE in PROF can only strengthen security in the PBS transaction supply chain, not weaken it.

\subsection{Mitigating Potential Attacks}\label{subsec:mitigating-attacks}
\subsubsection{(Grinding Attack) Leakage of information from the bid value}
The proposer's revenue, i.e., the PROF-enriched bid $b$, is comprised of the transaction fee from the PROF bundle, which in turn depends on the execution behavior of the PROF transactions during the simulation inside the bundle merger.
This execution of PROF bundle can very well depend directly on the contents of the prefix block $\pbsblock$ input to the bundle merger.
For example, multiple simulations with different input prefix blocks containing carefully crafted DEX (Decentralized Exchange) trades could reveal whether the PROF bundle contains a DEX trade for the particular asset pair and the details thereof.
This is because the transaction fees, or even validity of a DEX trade, depends on the trades executed before it.
In general, an adversary can learn useful bits of information about the PROF transactions by studying the relationship between chosen input blocks $\pbsblock$ to the bundle merger, and the output bids $b$.
In the above example, information obtained by the adversary can be exploited to frontrun the PROF users' DEX trades by inserting carefully crafted frontrunning transactions in the winning PBS block.

To mitigate leakage through this channel to the maximum extent possible, we require the following:
\begin{requirement}[Enrich only once]\label{req:enrich-only-once}
The PROF bundle is merged with only one prefix block ever -- the winning block $\winningpbs$ from the PBS auction -- so that only one PROF-enriched bid is available to an adversary.
\end{requirement}
 
Often in practice, this privacy concern is limited to the scenario wherein the adversary can use this leaked information to frontrun the PROF users through inserting transactions in $\winningpbs$. Here, the above requirement can be relaxed to the following:
\begin{requirement}[Begin enriching before revealing]\label{req:relaxed}
    The PROF bundle merger accepts no new prefix blocks for merging after it reveals a PROF-enriched bid. More formally, if the bundle merger reveals a PROF-enriched bid at time $T+\delta$ (for a prefix block input at time $T$), then it only accepts prefix block inputs until time $T+\delta$.
\end{requirement}
The relaxed Requirement~\ref{req:relaxed}, compared to the requirement~\ref{req:enrich-only-once}, allows for any late (and higher value) PBS blocks arriving between $T$ and $T+\delta$ to also be enriched with the PROF bundle, thus increasing the inclusion likelihood of the PROF bundle (Section~\ref{section:fee-latency}).
An astute reader may notice that this relaxation can potentially open the door to spamming of the bundle merger with different prefix blocks in an attempt to blindly frontrun the PROF transactions. 
While the window for spamming the bundle merger is fairly small (between $T$ and $T+\delta$), these blind attempts can be easily mitigated by placing a cap on the number of blocks enriched by the PROF bundle merger for any one slot.




\subsubsection{Replay Attacks on TEE}
As we discussed above, PROF requires that one or a limited number of PROF-enriched bids be revealed, to limit the bits of information revealed about the PROF transactions.
We therefore have to mitigate any rebooting and replay attacks on the PROF TEE, that bypass this requirement.
We store the PROF bundle, and an execution flag, inside the volatile memory, so that any PROF bundles received are no longer available after a reboot.
Additionally, for thwarting any replay attacks, the TEE simply waits for the inter-block interval (12 seconds in Ethereum) before starting to produce PROF-enriched bids. This ensures that even if an adversary is able to obtain additional PROF-enriched bids by means of replay, they are not going to be accepted by the blockchain network.

\section{PROF-Share: An Enhanced Design}
\label{section:double-auction-design}
\begin{figure}
    \centering
    \includegraphics[width=.99\columnwidth]{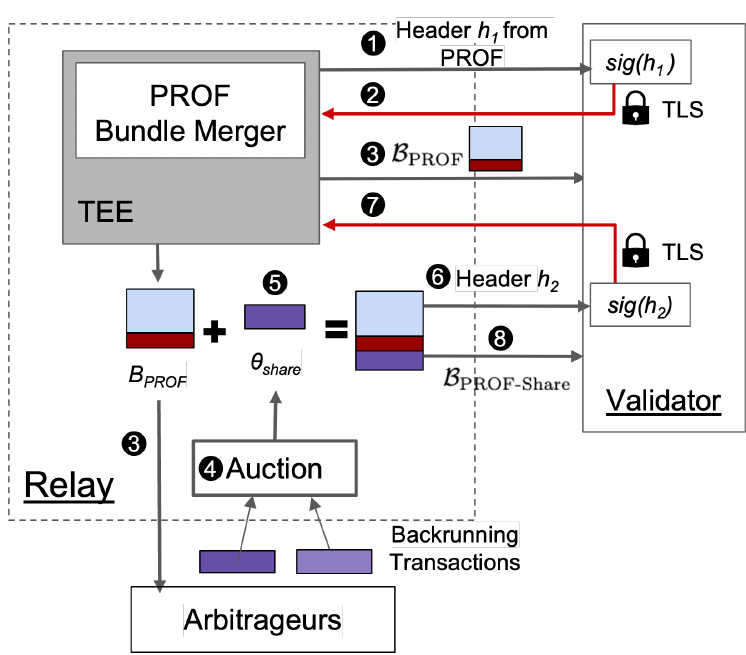}
    \caption{PROF-Share design. Steps 1-3 are the same as PROF. Once committed to by the validator, PROF-enriched block $\profblock$ is also released to arbitrageurs, who submit backrunning transactions in a second auction (step 4) and compete on providing PROF users with the most amount of kickback. The winning backrunning transactions are merged into the PROF-Share block $\profshareblock$, which is supplied to the validator after the validator commits to it (steps 5-8). 
    }
    \label{fig:profsharedesign}
\end{figure}




We now outline a \textit{backrunning auction design}, that we call PROF-Share.
It is an enhancement to the PROF design (Section~\ref{section:single-auction-design}), inspired by MEV-Share~\cite{mevsharedocs}.
It aims to further improve the utility for PROF users and blockchain validators by compensating them using profits obtained by backrunning the PROF bundle.
Unlike MEV-Share that allows backrunning of individual user transactions, PROF-Share only allows the backrunning of the PROF bundle as a whole. As we describe in detail in Section~\ref{sec:analysis}, this leads to higher redistribution of profits to the users and validators.

Figure~\ref{fig:profsharedesign} contains an overview of the PROF-Share design.
PROF-Share proceeds in the same way as the vanilla PROF design until the point at which the proposer sends the signed header to the PROF bundle merger. 

Unlike PROF, the bundle merger in PROF-Share does not publish the signed header from the proposer, but rather stores it for later.
It only releases the content of the underlying block from PROF ($\profblock$) publicly to arbitrageurs, builders, and searchers, who create backrunning transactions ($\sharebundle$) for the PROF bundle, to be appended to the end of the block. 

In other words, the final sequence of transactions in block $\profshareblock$ produced from PROF-Share is concatenation of transactions in $\profblock$ and $\sharebundle$.

Notice that at this point the proposer has already committed to the block $\profblock$ that contains the PROF bundle, but this commitment is stored exclusively inside the bundle merger. The sequence of transactions up to and including the PROF bundle is in a sense already finalized, preventing exploitation of the PROF transactions. 

The backrunning set of transactions $\sharebundle$ is chosen through another first-price auction at the relay.
More specifically, $\sharebundle$ that maximizes the users' revenue is selected by the relay, as long as the proposer also get some non-zero portion as an incentive to include $\sharebundle$.
Similar to the original auction in PROF, this second auction is also conducted outside the bundle merger.

Proposers can query the header for $\profshareblock$ along with the total reward, and return the signed header to the bundle merger via a secure TLS connection.
If the bundle merger obtains a valid signed header for $\profshareblock$ from the validator, it releases this signature and the auction module releases the contents of the block $\profshareblock$, which is then finalized by the blockchain network.
If the bundle merger does not obtain a second signature from the proposer before the end of the slot, it releases the previously stored signature for $\profblock$ produced using vanilla PROF, to be finalized by the blockchain network.
Note that the PROF bundle merger ensures that only one signed header is ever releases for each slot.

\mypara{Threat Model.}
As mentioned in Section \ref{section:threat-model-single}, our aim is to not require any more trust in the relay operator than is already tolerated by the network. 
The PROF-Share design requires the proposer to potentially sign two block headers for the same slot.
While our design must still allow for the proposer to be slashed if they indeed attempt to equivocate, an honest proposer following the PROF-Share protocol should not be unfairly slashed for following the protocol. Therefore the PROF bundle merger, running in the TEE, is responsible for ensuring that it releases only signed header, corresponding to either $\profblock$ or $\profshareblock$.
Similar mitigation as in Section~\ref{subsec:mitigating-attacks} are applied to prevent replay attacks on the TEE.



\section{Economic Analysis}
\label{sec:analysis}

We analyze PROF for the benefits it provides to the transaction senders, i.e., the users of the system.

In Section~\ref{sec:analysis-economics}, we analyze the economic utility $U$ to the users of PROF, PROF-Share and MEV-Share. Utility $U_p(S, \block)$ of block $\block$ at state $S$ for player $p$ is simply defined as the following:
\begin{equation}
   U_p(S,\block) = \textsf{bal}(p,S') - \textsf{bal}(p,S),
   \label{eq:utility}
\end{equation}
where $S \xrightarrow{\block} S'$ is a valid blockchain state transition and $\textsf{bal}$ denotes the balance of the player in a particular state.
To the best of our knowledge, we are the first to perform any systematic analysis of MEV-Share.
We limit our analysis of utility to a model where users are only sending AMM trades, as they form the bulk of known MEV activity. Since PROF protects users from frontrunning exploitation, including their transactions via PROF clearly leads to better utility than if the same transactions were sent unprotected and allowed to be frontrun.
On the other hand, mechanisms such as MEV-Share not only protect individual users from frontrunning, but give them kickbacks from the arbitrage profits obtained via backrunning their transactions. 
Therefore, we ask the following questions:
\begin{enumerate}[leftmargin=*]
    \item How does the average utility for a user compare across PROF, PROF-Share, and MEV-Share?
    \item Despite not giving kickbacks to users, are there market conditions under which PROF provides higher average utility than the redistributive mechanism of MEV-Share?
\end{enumerate}

\subsection{Economic Utility for PROF Users}\label{sec:analysis-economics}

We aim to understand how PROF and PROF-Share perform against MEV-Share (Section~\ref{section:mevsharebackground}) in terms of the economic utility defined in~\Cref{eq:utility}, in a model where users are only making AMM trades. 
MEV-Share protects users from frontrunning and shares profits derived from backrunning opportunities created by their trades.
Other redistributive mechanisms, such as MEV-blocker~\footnote{\url{https://mevblocker.io/}}, are equivalent to MEV-Share for the purposes of this work, and we refer to all these redistributive mechanisms as MEV-Share.
For our analysis, we create a simple model to simulate user trades and arbitrage opportunities created by user transactions, and then compare the average utility for users across the three mechanisms under various market conditions.
We look to traditional finance literature, specifically works on frequent batch auctions by Budish et. al. \cite{budish2015hftarm,budish2024market} to guide how we model user trades(Section~\ref{sec:demand-ratio}), and arbitrage behaviour. We chose this work as it focuses on applying the concept of batch auctions to traditional markets, which corresponds to the batching of blockchain transactions in a block.

\subsubsection{Setup}
In our model, users trade one of two tokens (Token $X$ or Token $Y$) on a single constant product AMM exchange. Trades are executed according to the constant product rule such that the product of liquidity ($L_{X}$ for Token $X$ and $L_Y$ for Token $Y$) before and after a trade remains the same. For example, if the trade is selling $\text{amt}_Y$ of Token $Y$ to get $\text{amt}_X$ of Token $X$ then $L_X\times L_Y = (L_{X}-\text{amt}_X)\times (L_Y + \text{amt}_Y)$. 


In our simulations, the exchange starts off with an initial state $S$ having the same liquidity of $L_{X}=L_Y=1e7$ for both tokens, and thus a 1:1 exchange price. Note that changing the initial liquidity ratio is only a matter of normalization, and would not affect the results qualitatively. All users start off with 100 units of both tokens, representing a capital that is neither too small to not discriminate the mechanisms, nor too large to generate unrealistic trade size relative to the pool's liquidity. Users independently choose to trade all of their Token $X$ or all of their Token $Y$ with equal probability. These trades are executed at the market price without any slippage limit.

As the transactions of a block $\block$ are executed in a particular order, the price obtained by a user transaction $t$ depends on the set of transactions preceding it. Clearly, the utility $U$ of a user depends on the block $\block$ produced by the mechanism.
These price fluctuations create arbitrage opportunities between exchanges. We model these arbitrage opportunities by introducing another exchange for the same two tokens, but one that has a static price, or equivalently, infinite liquidity. This second exchange mimics an external price oracle or a centralized exchange.
We set the price $P_{static}$ of this second exchange to be the same as the top-of-the-block price for our first exchange i.e. 1:1 exchange price.
In other words, we start from an initial state which has no arbitrage opportunity available.

Similar to~\cite{budish2015hftarm, budish2024market}, arbitrageurs in our model do not have any inherent preference for a particular asset: they perform the backrunning transactions by buying the cheaper token on one exchange and selling it back to the other exchange for a profit.
For example, if a user sold Token $Y$ to the first exchange the relative price of Token $X$ would increase. The arbitrager would then perform a \textit{backrun} by buying Token $X$ for the lower price on the second exchange and sell it to the first exchange at the higher price. To maximize this profit the arbitrager must buy enough amount $arb_X$ of Token $X$ such that its price would be equal on both exchanges. Thus, $arb_X = \sqrt{L_X\times L_Y \times P_{static}} - L_X$.

This model allows us to compare how effective PROF, PROF-Share, and MEV-Share are at allowing arbitragers to capture these opportunities and how this affects users' utilities in light of redistribution performed in the latter two systems.
In MEV-Share, arbitragers are allowed to perform backrunning after every user transaction. They then return 90\% (the default value in MEV-Share) of their profit as a kickback to the particular individual user. 
However, in PROF-Share, the user transactions are kept private, until they are committed as a bundle. Therefore, the arbitragers are only allowed to perform backrunning for the bundle as a whole, and any backrunning transactions are sequenced after the PROF bundle. 
The 90\% of the backrunning profit is split among all the PROF-Share users, in proportion of the size of their trades (our analysis of average utility is agnostic to how this split is carried out).
Finally, in PROF, no backrunning and thus, no redistribution by arbitragers is allowed.

\begin{figure} \includegraphics[width=\columnwidth]{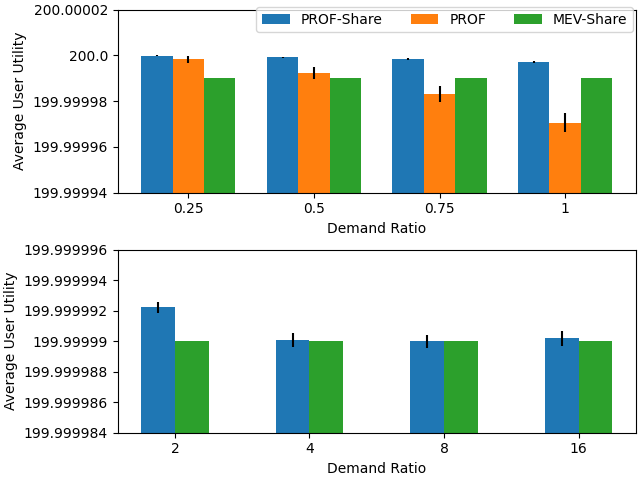}
    
    \caption{Average User Utility for Demand Ratio of 0.25, 0.50, 0.75, and 1, 2, 4, 8 for 20 to 100 users/block over 1,000 iterations. Error bars indicate the standard deviation. We do not plot the average utility for PROF for higher Demand Ratios, as it becomes negligible in comparison to redistributive mechanisms.}
    \label{fig:analysis}
\end{figure}

\subsubsection{Models of User Demand} \label{sec:demand-ratio}

To compare the performance of MEV-Share and PROF mechanisms, we examine user utilities in different market conditions. We characterize the market conditions using two parameters.

The first parameter, \textit{net demand}, is used by Budish et al.~\cite{budish2015hftarm} as an indicator of arbitrage opportunities in batch auctions. It is calculated as the absolute difference between the volume of trades in one direction and the volume of trades in the opposite direction within a batch. This value indicates the total price movement after the batch of transactions is executed. Similar to batch auctions, the arbitrage in PROF-Share is available only at the end of the PROF bundle, making net demand a good estimate of available arbitrage. 
Similarly, the redistribution opportunity left on the table by the basic design of PROF is dictated by the net demand parameter. Intuitively, a larger net demand implies lesser utility for users in PROF.
For a set of transactions $\text{tx}_{\block} = \{\text{amtX}_0, \ldots , \text{amtX}_n\}$ in a block $\block$ that trade on the same exchange, where \( \text{amtX}_i > 0 \) represents a trade where a user sells Token $X$ and \( \text{amtX}_i < 0 \) represents a trade where a user buys Token $X$, the net demand $D_\block$ in block $\block$, denominated in terms of Token $X$, is the absolute value of the sum of the transaction amounts $\text{amtX}_i$:

\begin{equation}
D_{\block} = \Big| \sum_{\text{amtX}_i \in \text{tx}_{\block}} \text{amtX}_i \Big|.
\end{equation}

The second parameter, \textit{variability}, represents the price movement between individual transactions. We calculate variability as the standard deviation of transaction volumes. In MEV-Share, arbitrage can occur after each transaction, making variability a good estimate of available arbitrage. For a set of transactions $\text{txs} = \{\text{amtX}_0, ... \text{amtX}_n\}$ where $\mu$ is the average volume of the transactions, the variability is the standard deviation of transaction \textit{volume} or the standard deviation of the absolute value of the transaction amounts,
\begin{equation}
 V =\sqrt{\frac{\sum_{i=0}^n (|\text{amtX}_i| - \mu)^2 }{n}},
\end{equation}
where $\mu$ is the mean of volumes $\text{amtX}_i$.

In~\cite{budish2015hftarm}, batch times are chosen according to a parameter ``maximum net demand'', that represents a cap on the  net demand in a batch.
In contrast, in the decentralized finance (DeFi) world, batch times are fixed to one block and cannot be adjusted to optimize PROF, PROF-Share, or MEV-Share. Therefore, we use simulation experiments to evaluate the average utility in different mechanisms under varying market conditions. For this purpose we define the \textit{demand ratio} as a ratio of our two parameters, net demand to variability,

\begin{equation}
    \text{Demand Ratio}=\frac{D_\block}{V}.
\end{equation}
Intuitively, a higher demand ratio reflects a higher propensity to trade in the same direction, thus creating more arbitrage opportunities at the end of the batch relative to the arbitrage opportunity after each trade.

For a given number of $N$ users per block on average, we randomly sample a sequence of trades (to either sell 100 Token $X$ or 100 Token $Y$). We run our simulation for varying upper bounds on the demand ratio.
During sampling, we ensure that the demand ratio of this sequence of trades is less than our desired upper bound.
As in~\cite{budish2015hftarm}, the number of transactions in a block is sampled through a Poisson process. Since the transaction amounts are the same, the volume also follows a Poisson distribution, and its standard deviation is simply $10 \sqrt{N} (= \sqrt{100N})$.

\subsubsection{Results and Interpretations} Figure~\ref{fig:analysis} shows the average utility in PROF and PROF-Share benchmarked against MEV-Share for various demand ratio parameters up to 1. We repeat our simulations for 20 to 100 average users per block (i.e. the mean of the Poisson process for volume) for 1,000 iterations each.

Our experiments show two important results: 
\begin{enumerate}
    \item PROF-Share always gives the highest average utility in our experiments, compared to other two mechanisms.
    \item For market conditions where the net excess demand is lower relative to variability (i.e. smaller demand ratio), PROF (without any redistribution) provides higher average utility than a redistribution mechanism such as MEV-Share.
\end{enumerate}
 
The intuition for the first result is as follows: 
In MEV-Share, a portion of the value is paid to the arbitrageur after each individual transaction. However, in PROF-Share, this value is often captured internally within the PROF bundle, such that only the \textit{residual} net demand is available for the arbitrageur to capture a portion of. More formally, as Remark~\ref{rmk:mev-share-constant} elaborates, the price received by each user in MEV-Share is $P_{static}$. On the other hand, consider a user $u$ in PROF-Share that trades in  direction $d$. Suppose that out of the user transactions preceding it, an $m$-fraction of them trade in the same direction $d$, and $n$-fraction trade in the opposite direction of $d$. Whenever, $m < n$, user $u$ gets a better price than $P_{static}$. Note that the order of trades preceding $u$ does not influence the price that $u$ gets, due to the path independence of constant product AMMs~\cite{buterin2017path}. In a sense, $u$ is inadvertently backrunning all the users that came before it, and consequently, capturing value that would have otherwise leaked to an external arbitrager. So PROF-Share gives higher average utility than MEV-Share. PROF-Share also clearly gives higher average utility than PROF, which lacks backrunning redistribution. 

The intuition for the second result is as follows:
PROF (without any redistribution) can provide a higher utility than MEV-Share for reasons similar to those above. PROF internalizes the value capture that would have otherwise leaked to an arbitrageur. In other words, value can flow from one user to the next user in PROF, if they trade in opposite directions. On the other hand, PROF does leave out the backrunning opportunity at the end of the PROF bundle, whereas MEV-Share exploits every backrunning opportunity and redistributes most of the profits back to the users.
Recall that the net demand directly dictates the available arbitrage opportunity after the PROF bundle. Therefore, with lower net excess demand, there is less redistribution opportunity left on the table at the end of a PROF bundle. 
We find that for period of low net excess demand (e.g., below 50\%), the internal value capture in PROF more than compensates for any redistribution value left on the table at the end of the PROF bundle. 
We also find that for even tighter limits on net excess demand (e.g., below 0.25), the utility of PROF approaches that of PROF-Share, as the PROF bundle does not leave any arbitrage (and consequently, kickback) opportunity on the table.
For higher values of max net demand (0.5 and 1) MEV-Share begins to deliver higher value than PROF, due to higher arbitrage and kickback opportunities available than any internalization of value in the PROF bundle. But again, MEV-Share achieves lower average utility than PROF-Share. 

\begin{remark}\label{rmk:mev-share-constant}
Note that in our model, the utility for every user in MEV-Share is identical and remains the same irrespective of the way user trades are generated. After every user transaction, arbitragers bring the AMM back to its original state by arbitraging against the external static market. Therefore, every user gets the same price $P_{static}$ as the external market, and also gets an identical kickback from the arbitrageur. However, if the external market is instead assumed to have finite liquidity, then the utility of users in MEV-Share also would depend the market conditions of the AMM trades.
\end{remark}


\subsubsection{Validation of Our Model using Real World Data} 

We calculated the demand ratio for the most popular real-world AMMs using historical blockchain data to gauge the benefits of using PROF in a realistic setting. 

We collected the addresses of all UniswapV3\footnote{\url{https://blog.uniswap.org/uniswap-v3}} and SushiSwap\footnote{\url{https://www.sushi.com/swap}} pools by collecting all \verb|PoolCreated| events on their respective factory contracts. We also collected all \verb|Swap| events, which are emitted every time the \verb|swap| function is called on a \verb|UniswapV3Pool| contract. The event contains the amount for each token being swapped and the pool address. We only kept the events that originated from UniswapV3 or SushiSwap contract addresses.

As described in Section \ref{sec:demand-ratio}, the demand ratio is the net demand in a batch normalized relative to the variability, which is standard deviation of the batch volume. Recall that the net demand is simply the absolute value of the algebraic sum of each swap amount in the given block.

Since a block may not have enough trades within itself for a reliable computation of the variability for each block, we compute the variability by also including trades in adjacent blocks. More precisely, to compute the variability of Block $\block$, we include all the swaps within a window of some $s$ swaps, centered around block $\block$.
Specifically, for a block containing $n$ swaps, where $\text{amtX}_0$ is the first swap on the pool in the block, $\text{amtX}_{-1}$ is the last swap on the previous block, $\text{amtX}_{n+1}$ is the first swap on the next block, and so on, and $\mu$ is the average volume of the transactions, we calculate the variability using the standard deviation formula\footnote{If $n > s$, i.e., if a block $\block$ contains enough data points, then we compute the variability $V_n$ by taking the standard deviation of volume across all the swaps in $\block$. We omit this condition from Formula~\ref{eq:variability} for visual clarity.},
\begin{equation}
\label{eq:variability}
 V_s =\sqrt{\frac{\sum_{i=(n-s)/2}^{ (n+s)/2} (|\text{amtX}_i| - \mu)^2 }{s}}.
\end{equation}
Our results are plotted for $s = 500$, but we obtain similar characteristics for a reasonable choice of $s$.

\begin{figure} \includegraphics[width=\columnwidth]{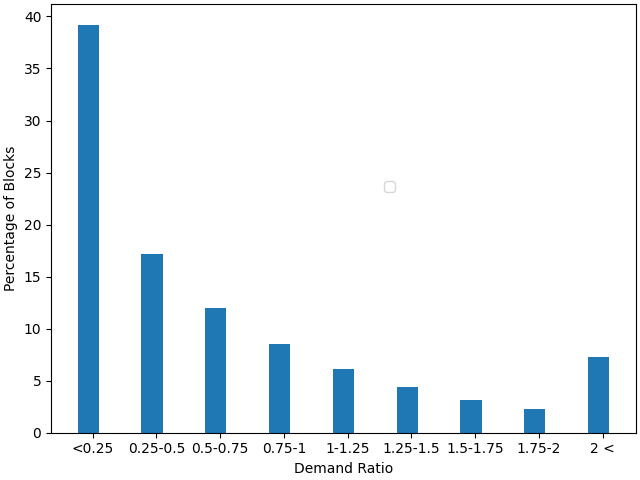}
    \centering
    \caption{Histogram of percentage of blocks falling into various demand ratio ranges. The percentage are calculated for each AMM pool separately, and then averaged across the pools.
    Variability calculated over a swap window of size $s$ = 500.}
    \label{fig:unfiltered}
\end{figure}

For a set of all blocks $\mathfrak{B}_z$ that contain swaps for a given AMM pool $z$, and a given range $x$ of the demand ratio,  we calculate the percentage, denoted by $y (x,\mathfrak{B}_z)$, of blocks that have a demand ratio in the range $x$.
In Figure~\ref{fig:unfiltered}, we plot the average percentage for each demand ratio range, using $V_{500}$ for the variability metric, averaged over a set of pools.

We find that the majority blocks have a demand ratio of $<0.5$, which is the market condition in which even the basic design of PROF delivers higher utility than MEV-Share.

\section{Implementation, Latency Analysis, and Evaluation}
\label{sec:impl}
All of our source code is open-source and can be found at \url{\sourceurl}. We use Kurtosis\footnote{\url{https://www.kurtosis.com/}} and our fork of its Ethereum package\footnote{\url{\sourceurl/ethereum-package}} to set up our development network and manage all entities involved, such as the PBS relay, builders, validator clients etc. 

\subsection{PROF-enabled Relay}
We start from a fork of the Flashbots' MEV-Boost Relay\footnote{\url{\sourceurl/prof-relay}} and modify the ``API'' component. 
A new HTTP endpoint is added to listen to and store the latest encrypted PROF bundles from the PROF sequencer in a Redis database. 
When the proposer sends the \texttt{getHeader} request, the builder's block with the highest bid, along with the latest PROF bundle is sent to a new component, bundle merger. 
The bundle merger is implemented using a fork of the Flashbots Builder\footnote{\url{\sourceurl/builder}} and we run it inside an Intel TDX Trust Domain\footnote{\url{https://github.com/canonical/tdx}}. Note that running the bundle merger inside the TEE only gives us conservative benchmarks of latency, compared to a benchmark obtained outside the TEE. 
Note that in order to create the block header, the bundle merger recalculates a new block hash, transaction root, state root, transaction receipts and logs, in accordance with the state changes obtained by applying the PROF bundle.
Any concurrent requests for new submissions from builders for the MEV auction continues to be processed by spawning new threads from the web-server.

\subsection{Analysis of the Effect of Latency on Inclusion}\label{section:fee-latency}
\begin{figure*}
    \centering
    \includegraphics[width=0.8\linewidth]{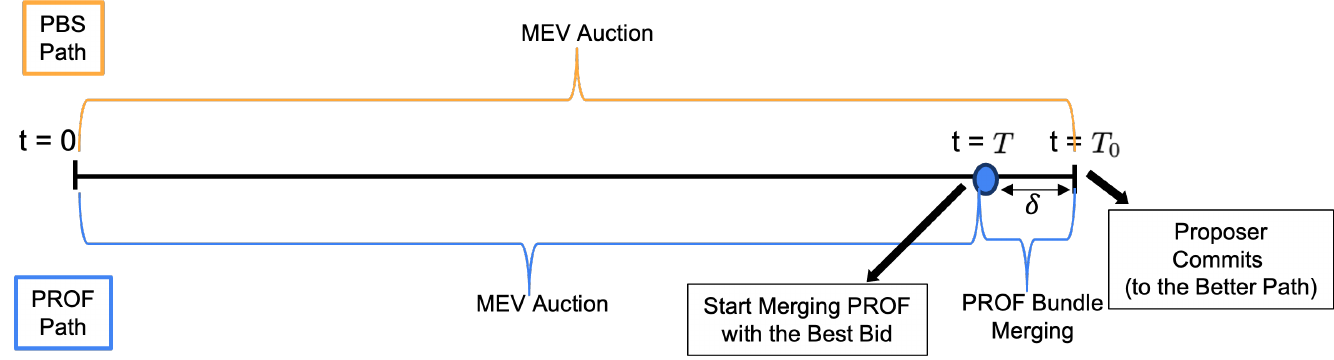}
    \caption{PROF timeline. The PBS path and PROF path are taken concurrently. $T$ is the parameter at which PROF bundle starts merging with the best bid in the MEV auction. Choice of $T$ is based on the latency of the implementation ($\delta$). Timeline not drawn to scale ($\delta \ll T_0$).
    }
    \label{fig:timeline}
\end{figure*}
We analyze the likelihood of inclusion of PROF transactions in the next block.
As we discuss below, it turns out that this likelihood varies directly with the transaction fees paid by PROF transactions, and varies inversely with the latency of securely simulating PROF transactions. Thus a key empirical question is:
\smallskip
\begin{itemize}
\item Can PROF transactions be included with high likelihood in the next block with minimal transaction fees paid and any reasonable latency?
\end{itemize}
\smallskip
To address this question quantitatively, we perform a measurement study on real-world data consisting of bids and submission timestamps in PBS auctions. Our analysis can be reproduced using the data and scripts at \url{\sourceurl/analysis}

Recall that the production of a PROF-enriched block happens concurrently with an MEV auction: The better of the two is made available to the proposer.
Therefore, the latency in simulating and appending a PROF bundle always has a neutral or positive impact on proposer revenue (Validator Incentive-Compatibility).
This small latency, however, may affect how competitive PROF-enriched block is compared to the winning block from the MEV auction, thus negatively affecting the inclusion rate of PROF transactions.

Intuitively, if the latency for appending a PROF bundle to a given prefix block $\earlyblock$ by the bundle merger is $\delta$, the winning builder's block $\lateblock$ in the non-PROF path has a $\delta$-duration period to improve over $\earlyblock$ enriched with the current PROF bundle.
For this PROF-enriched block to be selected by the proposer, the fee paid to the proposer in PROF transactions should at least make up the gap in bids of $\earlyblock$ and $\lateblock$ during the $\delta$ period at the end. Figure~\ref{fig:timeline} depicts the timeline of PROF.

We now express this break-even condition formally. Let $\mathsf{Bids}:\mathbb{R^+}\rightarrow\mathcal{P}(\mathbb{R^+})$ denote the stochastic process for the set of valid bids received by the relay up to a particular time instant\footnote{The set of bids over time is not monotonically increasing as relays allow bids to be canceled.}. 
Recall that the relay starts merging the latest PROF bundle ($\profbundle$) at time $T$.
Let $T_0 (>T)$ represent the time at which the relay receives the request that the validator would eventually commit to.
The fees from the PROF bundle are given by a deterministic non-negative function $\mathsf{Fees} (\cdot)$. The likelihood of inclusion $\alpha$ of $\profbundle$ is then given by:

\begin{equation}
        \label{eq:break-even}
        \text{Pr[} \mathsf{Fees}(\profbundle) >  
        \text{max}(\mathsf{Bids}(T_0)) - \text{max}(\mathsf{Bids}(T)) \text{]}.
\end{equation}

Given latency $\delta$ for producing a PROF-enriched header, we have $T = T_0 - \delta$~\footnote{In practice, exact $T_0$ might not be known a-priori, and therefore $T_0$ has to estimated in order to set the time $T$ for merging the PROF bundle. Fortunately, there is little spread in $T_0$ as validators wait until the slot deadline (12 seconds in Ethereum) to commit to a header~\cite{schwarz2023time}.}. Thus:

\begin{equation}
    \label{eq:latency-break-even}
    \begin{aligned}
    \alpha = \text{Pr[} &\mathsf{Fees}(\profbundle) >  \\
        &\underbrace{\text{max}(\mathsf{Bids}(T_0)) - \text{max}(\mathsf{Bids}(T_0-\delta))}_\textsf{Latency Penalty($\delta$)} \text{]}.
    \end{aligned}
\end{equation}

\begin{figure}
    \centering
    \includegraphics[width=.99\columnwidth]{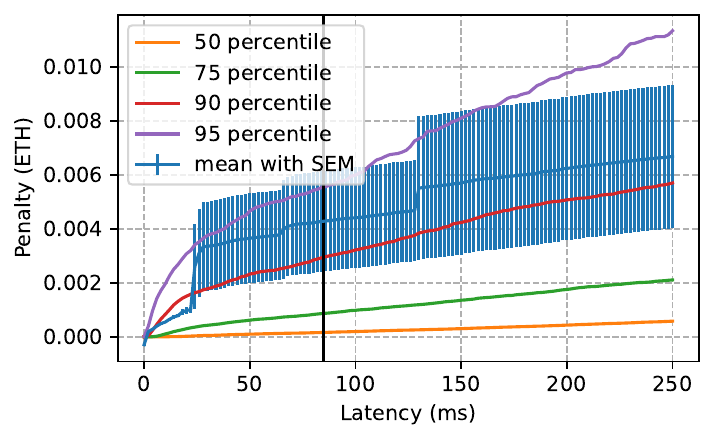}
    \caption{\textsf{Latency Penalty($\delta$)} characterized in a plot of penalty vs. total simulation latency, for 10,000 randomly selected historical slots (between Jan 3 - Apr. 11, 2024). The mean and percentiles are taken across these 10,000 slots. The solid vertical line at 85ms indicates the average latency of simulating a PROF-enriched block of size 15M gas (the current block size target in Ethereum).
    }
    \label{fig:latency-penalty}
\end{figure}

\noindent The term \textsf{Latency Penalty($\delta$)} represents the PBS value that a PROF-enriched block loses out on due to the latency in the PROF path at the relay. To characterize \textsf{Latency Penalty($\delta$)} in practice, we collect historical bids and their timestamp data from the Flashbots Relay API~\cite{flashbotsAPI} and plot the mean and percentiles of penalty over 10,000 slots randomly selected across 100 days (Jan. 3--Apr. 11, 2024) in Figure~\ref{fig:latency-penalty}. We also plot the penalty with data from large relays other than Flashbots' over the same period in Appendix~\ref{sec: other-relay}. They share a similar pattern, but have lower corresponding penalties.
Note that this data for bids is collected without deployment of PROF. It is well documented that builders are already engaging in late bidding behaviour, with bids rising disproportionately near the end (time $T_0$)~\cite{schwarz2023time}. Therefore, this data presents a conservative estimate of \textsf{Latency Penalty($\delta$)}: Once a relay enables PROF, we expect builders would send earlier bids (before time $T$) to increase their chances of merging with the PROF bundle.

As Figure~\ref{fig:latency-penalty} shows, for each percentile, the penalty approximately grows linearly with the PROF-bundle-simulation latency $\delta$.
For instance, to have a PROF-enriched block included at the Flashbots relay with a $90\%$ probability, the total fees from PROF transactions need to match the 90th-percentile penalty, which is roughly 0.022 ETH per second of simulation latency.



We now formulate the fee payment required \textit{per} PROF transaction in terms of the desired inclusion likelihood $\alpha$. Let the PROF bundle consume $\gasused$ gas and the fees per PROF transaction be $(1+\gamma) f$, where $f$ is the base fee, and $\gamma$ is the \textit{transaction-fee overhead} imposed by PROF (as a proportion of the base fees). Note that the base fee $f$ is mandatorily burnt, while the remainder $\gamma f$ goes to the validator.
Using Equation~\ref{eq:latency-break-even}, we have:

\begin{equation}
    \alpha = \text{Pr[} \gasused \gamma f >  \textsf{Latency Penalty($\delta$)} \text{]}.    
\end{equation}

As we observed earlier through Figure~\ref{fig:latency-penalty}, \textsf{Latency Penalty($\delta$)} is generally linear in latency $\delta$. Therefore, the above equation can be re-written as:

\begin{equation}
\label{eq:break-even-granular}
    \alpha = \text{Pr[} \gamma f >  \textsf{Latency Penalty}(\delta/\gasused) \text{]}.    
\end{equation}

The latency $\delta$ itself can be expressed as $\delta_0 + \gasused \beta$, where $\delta_0$ is a base latency overhead (minimum latency incurred for a PROF bundle of any size, due to database and network requests) and $\beta$ is the marginal latency for unit gas consumed (see Section~\ref{sec:eval-latency} for values of $\delta_0$ and $\beta$).

\begin{figure}
    \centering
    \includegraphics[width=.99\columnwidth]{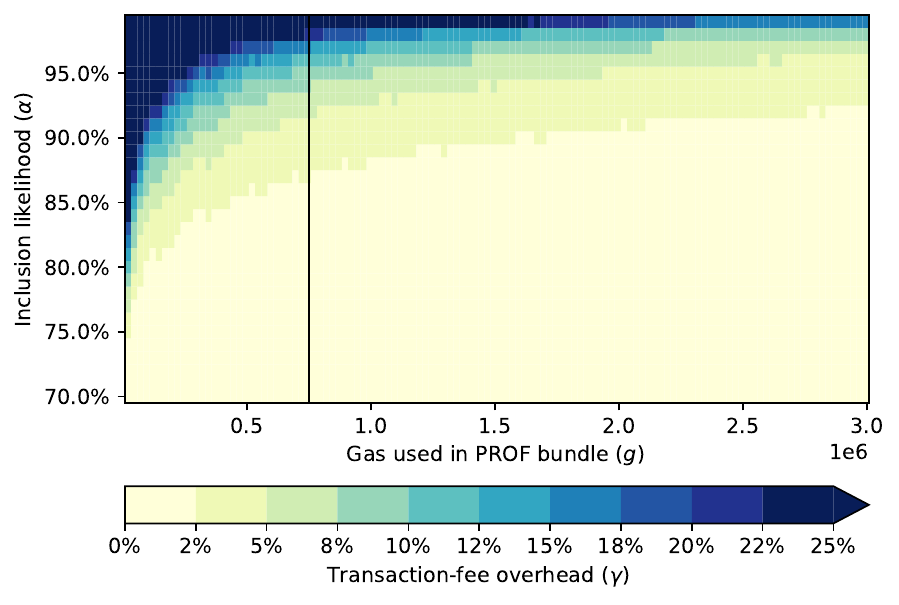}
    \caption{Transaction-fee overhead($\gamma$) vs. PROF-bundle inclusion rate ($\alpha$) vs. gas used in a PROF bundle ($\gasused$). The color of each point in the scatter plot indicates the range of transaction-fee overhead. For example, for a PROF bundle of size 750k gas (roughly 5 AMM swap transactions), to have an inclusion likelihood of $>$95\% for PROF-enriched blocks, charging 10\% more transaction fee will be sufficient.
    }
      \label{fig:multi-epoch-bench-other}
    \end{figure}

Using the same data set of $\mathsf{Bids}$ from Flashbots relay, and the base fee ($f$) data, we plot the three dimensional relationship among the inclusion likelihood ($\alpha$), the PROF \textit{transaction-fee overhead} ($\gamma$), and the gas consumed by PROF bundle ($\gasused$) in Figure~\ref{fig:multi-epoch-bench-other}. 
For instance, for a PROF bundle of size 750k gas (2.5\% of block capacity, and roughly 5 AMM swap transactions), a transaction-fee overhead of 10\% (i.e., $\gamma = .1$) will give a $>$95\% likelihood of inclusion in the next block.
The data for other large relays (Appendix~\ref{sec: other-relay}) indicates an even higher inclusion likelihood for the same transaction-fee overhead.

\subsection{Optimistic Relay and Merging at the Builder}
\label{sec:buildermerge}
\begin{table*}
    \centering
    \begin{tabular}{|c|c|c|c|c|c|}
        \hline
            & \textbf{Relay} & \textbf{Builders}  & \textbf{PROF Bundle Merger} & \textbf{Grinding Protection}  \\\hline
        PROF    & Pessimistic   & Completely untrusted  & Hosted at relays & \makecell{Both Requirement~\ref{req:enrich-only-once} \& \\ Requirement~\ref{req:relaxed} enforceable}   \\\hline
        PROF in hybrid mode & Optimistic    & \makecell{Trusted to produce \\ valid bids} & Hosted at builders & Only Requirement~\ref{req:relaxed} enforceable\\\hline
    \end{tabular}
    \caption{PROF can be deployed in hybrid mode to take advantage of the increasingly popular optimistic relaying method. This ensures that the Latency Penalty (formally defined in Section~\ref{section:fee-latency}), i.e., the loss from any higher late bids in the PBS auction, is much smaller for hybrid mode, resulting in higher inclusion likelihood for the PROF bundle. While builders cannot be prevented from enriching their blocks multiple times in the hybrid mode (Requirement~\ref{req:enrich-only-once}), grinding attacks are still prevented as the relay does not accept enriched blocks after it reveals a PROF-enriched bid (Requirement~\ref{req:relaxed}).
    }\label{tab:hybrid}
\end{table*}
Optimistic relays~\cite{optimistic-relay-doc} have grown in  popularity. Ultrasound Relay\footnote{\url{https://relay.ultrasound.money/}} introduced the notion of optimistic relaying, and is currently the dominant relay (delivering $>30\%$ of the blocks).
An optimistic relay dramatically reduces its latency by forgoing simulation of the builder's block in the critical path, and instead trusts (aided by collateral) the builders to produce a valid block and corresponding bid value. If the builder's block is found to be invalid after the fact, the relay could confiscate the collateral, and ban the builder from future auctions.

PROF too can take advantage of this optimistic method of relaying, by operating in what we call, a hybrid mode. 
In this hybrid mode, the PROF bundle merger TEE is hosted by builders. Since, builders are not trusted entities like relays, the use of TEE for housing the bundle merger is important for ensuring the privacy of PROF transactions in a hybrid deployment. 
The PROF sequencer sends the PROF bundle to the bundle merger TEE.
The bundle merger sends the PROF-enriched block, along with the associated bid (revenue to the proposer), to the optimistic relay over a secure channel. 
This ensures that builders learn nothing about the contents of the PROF transactions.
The relay now simply supplies the proposer with the block header for the best bid (out of all the PBS bids and PROF-enriched bids).

Unlike the vanilla PROF design, the bundle merger can now be used by builders to merge the PROF bundle with as many of their blocks as desired. 
However, the relay mitigates the leakage of the PROF transactions through the bid value (Section~\ref{subsec:mitigating-attacks}) in a similar fashion to the vanilla PROF design (Requirement~\ref{req:relaxed}) -- it does not accept any new bids after it has revealed a PROF-enriched bid (to the proposer).

We report in Section~\ref{sec:eval-latency} the latency of PROF in hybrid mode.
Table~\ref{tab:hybrid} summarizes the hybrid mode of PROF.


\subsection{Latency Evaluation}
\label{sec:eval-latency}

We run our evaluation inside an Intel TDX Trust Domain (TD), running as a virtual machine with 16GB RAM and on an Intel TDX machine with Intel Xeon Platinum 8570 processor. 
We evaluate the total additional latency introduced by PROF in the critical path of the \verb|getHeader| method of the PROF-enriched relay. This latency arises mainly in three parts: fetching the best block from the auction so far, simulating the PROF-enriched block inside the bundle merger, and signing the bid message to the validator. 

We find that there is a constant overhead $\delta_0$ of about 6.25ms regardless of the number of transactions. The marginal latency ($\beta$) of simulating a transaction is about 5.26ms per million gas, which amounts to $<$1ms for most transaction types (an AMM swap consumes roughly 150k gas). 
Note that while our benchmarks are based on a devnet which has a small blockchain state, the marginal transaction simulation latency on the mainnet state is similar~\cite{reth-paradigm}.
The constant overhead is independent of the size of the state.

\begin{figure}
    \centering
    \includegraphics[width=.99\columnwidth]{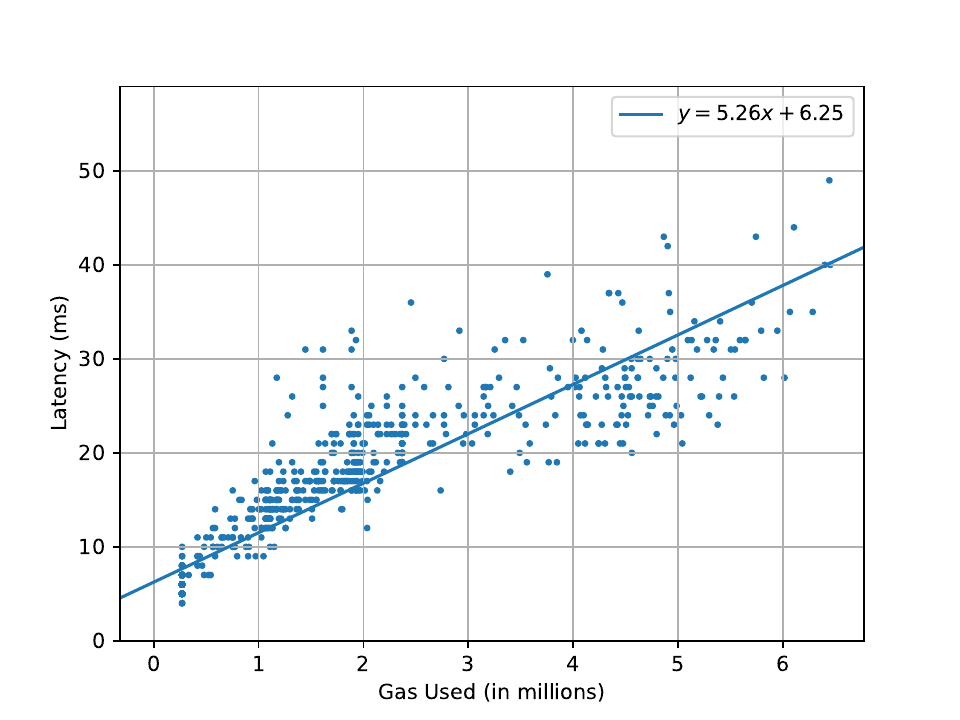}
    \caption{Additional Latency in the critical path vs. size of PROF bundle (in gas). For reference, an AMM swap transaction consumes roughly 150k gas, and the target block size on Ethereum is 15 million gas.}
    \label{fig:latency-benchmark}
\end{figure}

Figure~\ref{fig:latency-benchmark} shows the additional latency due to PROF in the critical path.
According to Section~\ref{section:fee-latency}, this latency is already sufficient to grant a high inclusion rate with minimal overhead in transaction fees for PROF users.

\section{Related Work}
\label{sec:related}
In recent years, blockchain transaction ordering has become an active research topic. 
Below, we highlight several works in this area and provide comparisons to PROF. 

\mypara{Ordering protocols.}
A substantial line of work builds protocols to provide specific guarantees on the finalized transaction ordering---causal ordering~\cite{cachin2001asyncbroadcast, malkhi2023mev} ensures that transaction contents are revealed only after their sequence is finalized; similarly, time-based ordering~\cite{kelkar2020order,zhang2020oligarchy} aims to order transactions by time of receipt. 
However, all such protocols require an honest majority---making instantiation challenging in a rational players setting.
As stated earlier, a primary goal for PROF is to maintain the strong ordering guarantees from these protocols while preserving incentive-compatibility (IC) for rational participants.

\mypara{Transaction fee mechanisms.}
\cite{roughgarden2021transaction,chung2023foundations} formalize how transaction fees need to be set to provide IC for both users and block-producers. However, Bahrani et al.~\cite{bahrani2023transaction} show a significant barrier---if block-producers attempt to actively extract MEV, as is the case in practice, no fee mechanism can be incentive compatible. The authors posit that the impossibility could be circumvented through the use of e.g., cryptography or trusted hardware---our approach in PROF provides one potential direction to accomplish exactly this. 

\mypara{Time and latency advantages.}
Seminal work by Daian et al.~\cite{daian2020flashboys}---which initiated investigation into MEV---highlighted how similar to traditional finance, timing advantages were significant within DeFi. It is a well-known folklore result that more MEV can be obtained by playing \textit{timing games}---waiting longer to e.g., include more transactions or to find more profitable orderings. 

Recent work~\cite{time-is-money} further studies timing games that could be played by proposers in PBS.
The time of initiating bundle merging in PROF (parameter $T$) can be set according to the timing preferences of the designated proposer for the current slot (after taking into account the latency of PROF bundle merger). If the proposer does not make its timing preference known to the relay, $T$ can be set with enough margin to tolerate any timing games. 
Real-world analysis in~\cite{time-is-money} shows that, at least currently, significant timing games are not being played (likely due to social or community pressures), and therefore it is straightforward to determine the time for initiating bundle merging in PROF.

\subsection{In-protocol mechanisms}

The Ethereum community has considered multiple proposals that seek to mitigate trust, centralization and security challenges that have arisen from out-of-protocol PBS.
All these proposals require sweeping changes to the core protocol, but we nevertheless discuss them below.

\mypara{Enshrined-PBS (ePBS).}
There have been several proposals \cite{EPBSInfinite, ethereumEIP7732Enshrined} aimed at decentralizing or removing centralized relays from PBS, by \textit{enshrining} PBS within the Ethereum protocol\footnote{Unfortunately, due to bypassability \cite{relaysPostePBS}, it is not possible to enforce that proposers utilize enshrined PBS mechanisms over an out-of-protocol mechanism---such as MEV-boost.
If an incentive exists to utilize an out-of-protocol mechanism,
rational proposers will indeed opt out of ePBS.}. 
Ethereum Foundation's ePBS proposal, EIP-7732 \cite{ethereumEIP7732Enshrined}, proposes replacing the relay with a decentralized committee, sampled from the set of Ethereum validators. 
The responsibility of this committee is reduced to only ensuring fair data exchange between builders and validators, while block validation is delayed to a future slot. 
This is similar to the role of optimistic relays in the hybrid deployment of PROF (Section~\ref{sec:buildermerge}).
Indeed, PROF in its hybrid mode is compatible with EIP-7732, but with one caveat: 
PROF would require a mechanism to enforce 
Requirement~\ref{req:relaxed} (i.e., PROF-enriched bids are not accepted after a PROF enriched bid is revealed) in order to mitigate against grinding attacks (Section~\ref{subsec:mitigating-attacks}).
We argue that similar mechanism would be needed by builders in EIP-7732 for privacy of their bids, which is currently provided by the relay in MEV-boost.


\mypara{PEPC.}
Protocol-enforced proposer commitments (PEPC) \cite{unbundlingPBS} describes an in-protocol mechanism that allows proposers to enter into \textit{arbitrary} commitments with builders over the blocks they build. 
For example, a proposer can enforce that the builder includes particular transactions, uses a particular ordering scheme, or allocates distinct segments for different optimization objectives.
While details of PEPC are complex and somewhat unclear, PROF represents one simple mechanism for PEPC that would provide protection for users as well as maximize value for rational validators.

\mypara{Inclusion Lists.} The centralization of builder market in PBS~\cite{yang2024decentralization} has created risks of censorship. As a mitigation, several \textit{Inclusion Lists} designs (e.g., FOCIL~\cite{focil}) have been proposed. Inclusion Lists designs attempt to enforce inclusion of certain transactions that might be at the risk of censorship, as long as some honest validator observes the concerned transactions.
PROF obviates the need of Inclusion Lists to the extent that transactions at the risk of censorship can be included through a PROF bundle at a PROF-enabled relay, without needing any support from the builders. 

\subsection{Distributed block building}
In this section, we discuss a number of protocol proposals that---with the aid of TEEs---enable various mutually-distrusting parties in the supply chain to cooperatively engage in distributed block building.
The majority of this line of work \cite{eigenlayerMEVBoostLivenessfirst, mevbootee, pepcboost} seeks to address concerns orthogonal to PROF, such as censorship and economic centralization; 
however, select concurrent work \cite{flashbotsMEVMSUAVE} seeks to mitigate MEV.


\mypara{SUAVE.} 
SUAVE \cite{flashbotsMEVMSUAVE} is a proposed decentralized platform for confidential computation, designed to coordinate block production between searchers, builders and relays. 
Most notably, the platform provides access to a network of confidential compute providers (\textit{kettles}), anchored in a public permissionless blockchain.
Kettles are responsible for serving compute requests in a TEE, based on the semantics and current state of SUAVE's virtual machine, the MEVM---a modified version of the EVM that provides precompiles for transaction simulation and data storage access. 
In Appendix~\ref{app:suave}, we describe how SUAVE can be used to implement PROF, in theory.
We note, however, that SUAVE is currently in active development and unfortunately, does not yet provide the confidentiality and integrity guarantees necessary to mitigate grinding attacks against protected transactions (see Section~\ref{subsec:mitigating-attacks}). 
Moreover, SUAVE’s generality may come at the cost of a larger TCB, as well as additional latency overheads (e.g., due to virtualization). 

\mypara{MEV-boost+/++.}
This proposal \cite{eigenlayerMEVBoostLivenessfirst} puts forward two potential modifications to MEV-boost---MEV-boost+ and MEV-boost++.
Both aim to reduce censorship concerns by allowing the \textit{proposer} to append a bundle of transactions to the end of the winning builder top-of-block. 
The threat of slashing prevents proposer equivocation as before; however, the proposer must now sign partial blocks rather than full blocks. 
MEV-boost++ has the added goal of reducing trust in the relay by replacing it with a generic data layer. 
Submission of invalid partial blocks or inaccurate bids to the data layer leads to slashing. 


\mypara{MEV BooTEE.} MEV BooTEE \cite{mevbootee} shares similar objectives. 
However, instead of relying on economic incentives, the protocol leverages a TEE to assemble blocks and execute the duties of the relay. 
Builders provide blocks, while the proposer provides a bundle to be appended to the end. 
The authors propose three protocol variants, distinguished by which party runs the TEE assembler---namely, the proposer, builder or a separate standalone entity, such as the relayer. 

Their "Builder-Aide" design most closely resembles PROF as the proposer only sends its transaction list to be assembled with the highest bidding block. 
In PROF, the goal is to maximize the proposers final bid; we therefore allow for late blocks with a higher bid than the block with the PROF bundle included to be returned to the proposer.

\mypara{PEPC-Boost.}
Like PROF, PEPC-Boost \cite{pepcboost} also utilizes the top-of-block / bottom-of-block paradigm, allowing builders to submit separate bids for each partition. 
Unlike PROF, its objective is to reduce centralization in the block building market. 
The top-of-block is reserved only for transactions that engage in CEX-DEX arbitrage, in order to preserve block building independence for the rest-of-block.



%

\section{Discussion}
\label{sec:discussion}
While PROF is agnostic to the choice of the ordering policy for transactions, we briefly discuss a key benefit bestowed by PROF to a FCFS (First-Come, First-Served) ordering policy. We also discuss considerations around the role of a relay as a perceived neutral entity in block building.

\mypara{Alleviating latency racing in FCFS policy through PROF.} FCFS / FIFO (First-In, First-Out) ordering policy is commonly used in centralized exchanges (e.g., NSE~\cite{nse-fcfs}) as a means of providing fair ordering for traders. A long line of literature on decentralized ``fair ordering'' protocols~\cite{cachin2001asyncbroadcast,malkhi2023mev,Kavousi2023blind,li2023transaction,li2023ratel} also utilize FCFS for ordering blockchain transactions.
However, a big negative externality remains: FCFS incentivizes an arms race in reducing latency to the centralized operator~\cite{budish2015hftarm} or the set of decentralized nodes~\cite{babel2022strategic,tang2023strategic}, in order to capture common but profitable opportunities, such as price arbitrage. Indeed, transactions at the top of the block in Ethereum are often used to capture arbitrage opportunities created by the stale prices from the previous block~\cite{gupta2023centralizing, heimbach2024non}. An FCFS policy in PROF, however, \textit{does not} promote latency racing, as the PROF bundle is only \textit{appended} to the PBS block. The top of the block-space is still constructed via the auction mechanism of PBS. In other words, auctioning off the prefix of the block through PBS and then ordering transactions through FCFS mitigates the key negative externality of latency racing.

\mypara{Neutrality of PBS Relay.}
PBS relays are usually seen by many practitioners as neutral entities that do not take part in building blocks, or choosing which transaction sequencers to partner with. We highlight that a PROF-enabled relay does not need to choose a particular PROF sequencer exclusively or prioritize one PROF sequencer over another. As described in Section~\ref{sec:design-considerations}, a PROF-enabled relay can take in PROF bundles from multiple PROF sequencers---possibly implementing different ordering policies---and then carry out PROF bundle merging independently and concurrently for each of the sequencer, and select the best out of all the PROF-enriched block.
The relay could also combine multiple PROF bundles into one single PROF-enriched block. 
\section{Conclusion}
In this work, we examine the emergence of PBS and the MEV landscape. 
Specifically, we explore the design space for incentive-compatible MEV-protection mechanisms.
We introduce PROF, a novel architecture that integrates seamlessly with PBS to protect user transactions from MEV. 
PROF does so by sequencing encrypted transactions at the end of a block, utilizing underused block space produced by EIP-1559. 
PROF improves fairness and efficiency, and redistributes MEV back to users---without breaking backwards-compatibility or introducing additional trust assumptions.
We analyze PROF’s utility, incentive-compatibility and inclusion rate, and realize the protocol in an end-to-end implementation, demonstrating its practicality.

\section*{Acknowledgments}
This work was funded by NSF CNS-2112751, the Sui Foundation, and generous support from IC3 industry partners and sponsors. We thank Intel for access to a TDX instance through the Software Development Platform for benchmarking. Andrew Miller is currently working with Flashbots, an organization developing prominent products related to MEV. Any opinions, findings, conclusions, or recommendations
expressed here are those of the authors and may not reflect those of these sponsors and organizations.


\bibliographystyle{plain}
\bibliography{main}

\ifSP
\appendices
\else
\appendix
\fi

\section{PROF on SUAVE}
\label{app:suave}


SUAVE is a platform under development by Flashbots that aims to create a ``market for (block-building) mechanisms,'' i.e., to allow MEV mechanisms to compete within an open marketplace for creating the most profitable block for the validator~\cite{flashbotsMEVMSUAVE}. SUAVE also aims to provide ``programmable privacy'' to transactions during the block building phase. While the architecture is still evolving, SUAVE is meant to allow a specific mechanism to be deployed in its own protected environment known as a \textit{kettle}. A kettle is a TEE-enabled node running an EVM enhanced for MEV applications and called the MEVM. (Applications here are known as \textit{SUAPPs}.) Current SUAVE protocol specifications are available at~\cite{SUAVEspecs:2024}. 

In theory, as SUAVE's MEVM allows for general-purpose computation, the PROF sequencer and bundle merger could be implemented as SUAVE smart contracts, i.e., SUAPPS~\cite{SUAVEspecs:2024,flashbotsMEVMSUAVE}. (Indeed,~\cite{flashbotsMEVMSUAVE} specifically mentions PROF.)
For example, a bundle-merging contract may take as input the winning block from a contract which runs an MEV auction for partial blocks, along with the bundle produced by another fair ordering sequencer contract; the merger contract then simulates the merged block's execution using an MEVM precompile, to compute and output the complete block with its header. 

Due to the generality of the framework, however, the analysis for ensuring privacy of transactions from malicious \textit{downstream} contracts, and malicious kettle operators is quite challenging. For example, in the current specification~\cite{SUAVEspecs:2024} that is guiding the development of SUAVE, it is unclear whether and how SUAVE provides protection against grinding attacks on private transactions---a critical requirement for the approach underpinning PROF.


\section{Effect of Latency at Other Large Relays}
\label{sec: other-relay}
\begin{figure}[ht]\centering
    \centering
    \subfloat[Penalty vs latency.]{\label{fig:latency-penalty-ultrasound}
    \includegraphics[page=1, width=\columnwidth]{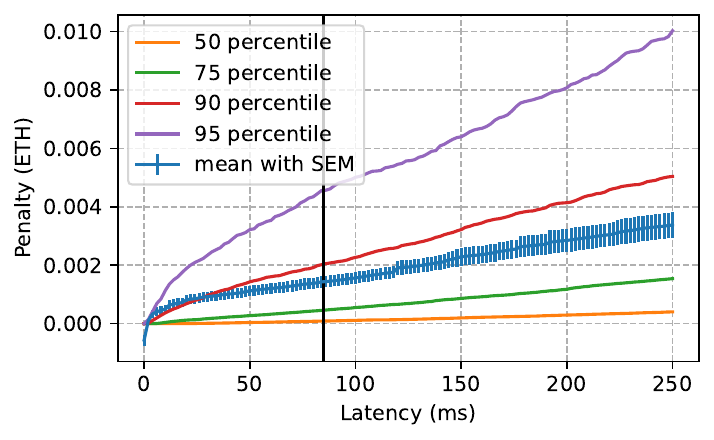}
    }
    \hfill
    \subfloat[Tx cost overhead vs inclusion rate.]{\label{fig:inclusionrate-feeoverhead-ultrasound}
    \includegraphics[page=1, width=\columnwidth]{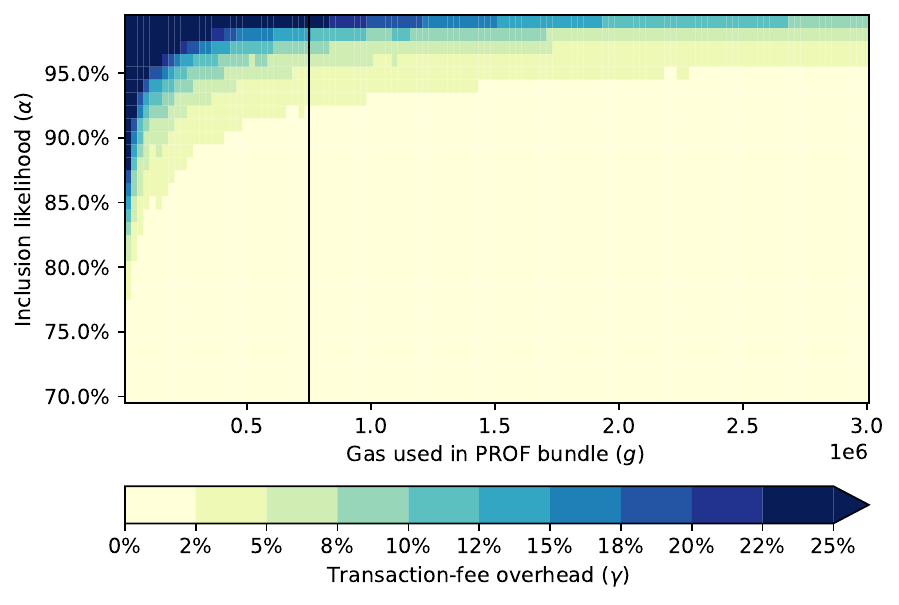}
    }
      \caption{Effect of latency according to Ultra Sound relay data.}
      \label{fig:multi-epoch-bench-other-ultrasound}
\end{figure}

\begin{figure}[ht]\centering
    \centering
    \subfloat[Penalty vs latency.]{\label{fig:latency-penalty-agnostic}
    \includegraphics[page=1, width=\columnwidth]{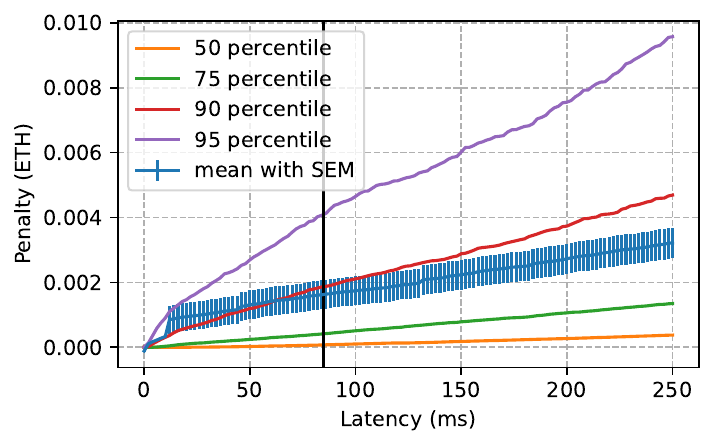}
    }
    \hfill
    \subfloat[Tx cost overhead vs inclusion rate.]{\label{fig:inclusionrate-feeoverhead-agnostic}
    \includegraphics[page=1, width=\columnwidth]{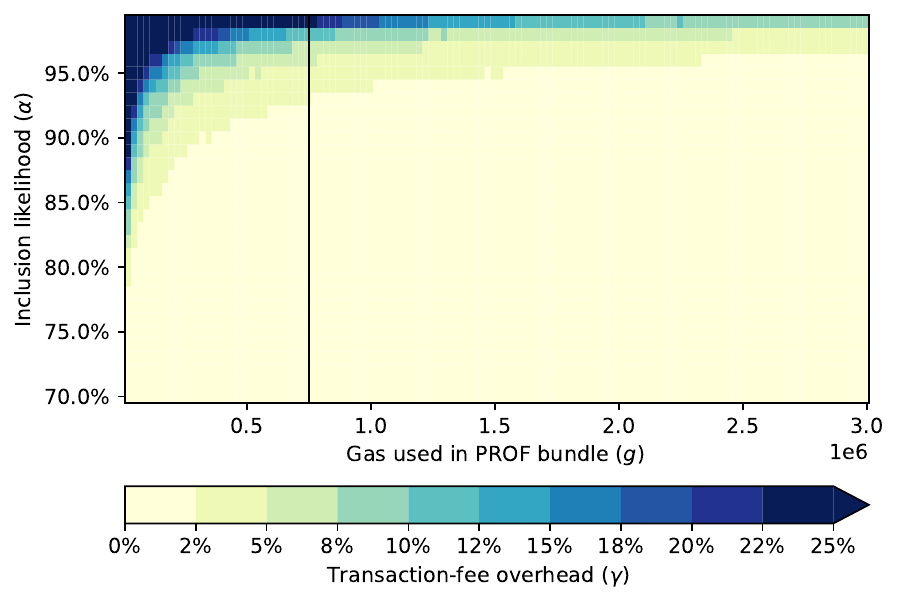}
    }
      \caption{Effect of latency according to Agnostic relay data.}
      \label{fig:multi-epoch-bench-other-agnostic}
\end{figure}

We plot the effect of latency with data from other large relays, namely Ultra Sound in~\Cref{fig:multi-epoch-bench-other-ultrasound} and Agnostic in~\Cref{fig:multi-epoch-bench-other-agnostic}.
These two relays and Flashbots relay together contributed $63\%$ of the blocks proposed during the period of which data is collected. 
The data for these two relays shows a slightly lower corresponding Latency Penalty than that in Flashbots relay (Figure~\ref{fig:latency-penalty}), which translates to a higher inclusion likelihood for PROF bundles (to the extent these relays also receive the winning bid of the PBS auction).

\end{document}